\documentclass[conference]{IEEEtran}
\usepackage{amsfonts}

\usepackage{amsmath}
\usepackage{centernot}

\usepackage{hyperref}

\ifCLASSINFOpdf
\else
\fi
\usepackage{mathtools}

\usepackage{graphicx}

\usepackage{multirow}

\begin{document}

%
\title{Toward Software Measurement and Quality Analysis of MARF and GIPSY Case Studies a Team
13 SOEN6611-S14 Project Report}


\author{\IEEEauthorblockN{Abdulrhman Albeladi}
\IEEEauthorblockA{Concordia University\\
Montreal, Canada\\
Email: blady911@gmail.com}
 
\and
\IEEEauthorblockN{Rabe Abdalkareem}
\IEEEauthorblockA{Concordia University\\
Montreal, Canada\\
Email: abdrabe@gmail.com}

\and
\IEEEauthorblockN{Farhat Agwaeten}
\IEEEauthorblockA{Concordia University\\
Montreal, Canada\\
Email: farhategweteen@gmail.com}
\and
\IEEEauthorblockN{Khalid Altoum}
\IEEEauthorblockA{Concordia University\\
Montreal, Canada\\
Email: toum.edu@gmail.com}
\and
\IEEEauthorblockN{Youssef Bennis}
\IEEEauthorblockA{Concordia University\\
Montreal, Canada\\
Email: youssef.bennis84@gmail.com}
\and
\IEEEauthorblockN{Zakaria Nasereldine}
\IEEEauthorblockA{Concordia University\\
Montreal, Canada\\
Email: zakaria.nasereldine@gmail.com}}


%


\maketitle

\begin{abstract}
It is no longer a debate that quality is an essential requirement in any software product, especially in a highly competitive market and a context of mission critical product. To obtain better product quality, software metrics are the only reliable indicators provided to assess and measure this attribute of a software product. Several metrics have been elaborated but none of them were really convenient in an object oriented ecosystem. However, the MOOD metrics have proven their efficiency in gauging the software quality at system level, while CK Metrics measure the quality of software at class level . These metrics, well suited for Object-Oriented design, allow measuring object oriented design properties such as coupling, cohesion, encapsulation, Inheritance and polymorphism. The goal of the present study is using the mentioned metrics to assess the quality of two different case studies, MARF and GIPSY. For this purpose, different tools such as McCabe, Logiscope and, JDeodorant have been used to measure the quality of these projects by implementing in different manners the metrics composing the CK and MOOD suite metrics, whilst MARFCAT has been used to detect vulnerable code files in both case studies. The present study puts the light on the strengths of these tools to measure the quality of proven and largely researched software products. 
\end{abstract}


%


\section{Introduction}
The size and complexity of software continue to grow throughout the development life cycle of a software system (Lehman’s Law), which in terms leads to an increase in the effort and cost of software maintenance [27]. These two software characteristics are considered  as internal quality attributes to the development team, which have a strong influence on external quality attributes such as functionality, usability, reliability, efficiency, reusability, extendibility, and maintainability [31] of a software system  during and  after its development. For this reason,  there is  a need  to have a methodology to measure or predict some internal attributes [28] such as size, complexity, functionality, reuse, modularity, coupling and cohesion, which may have direct impact  on the external quality attributes such as understandability, usability, reusability,maintainability. Theses internal and external quality provide an overview about the quality of the software product, and are dependent on its metrics. Therefore, Metrics have been considered as one of the means to obtain more reasonable assessments about milestones of any software project, or a software system developed that has lower faults [28].  Thus, there are many different metrics  such as  object oriented metrics,  which have been proposed by many researchers from different viewpoints to measure  the characteristics of  an object oriented      design [28] (such as  inheritance, polymorphism, information hiding and coupling). Good development teams can use the most important metrics to assess the quality of  design, and the structure of source code of  any software system from early stages  to keep track and control of the design of the software system to reduce complexity and  improve  the capability of the software system to adhere quality standards.

In this paper,  we will explore two different open source systems MARF ( Modular Audio Recognition Framework ) and GIPSY (The general intensional programming system) and their quality attributes. After that we will illustrate several types of metrics relevant to Object Oriented Design such as Chidamber and Kemere (C \& K) Metrics [32] , Metrics for Object Oriented Design (MOOD) metrics [17,20] , QMOOD metrics (Quality model for object-oriented design) [19,20] and other metrics.  The main objective of this study is to  focus on , four of C \& K Metrics such as Number Of Children (NOC), Depth of Inheritance Tree (DIT), Response for Class (RFC), Lack of Cohesion of Methods (LCOM) and six MOOD Metrics such as Method inheritance factor (MIF), Attribute Inheritance Factor (AIF), Method Hiding Factor (MHF), Attribute Hiding Factor (AHF), Coupling factor (CF), and  Polymorphism Factor (PF) to assess the quality of object oriented design for the both abovementioned open source systems, MARF and GIPSY. These ten metrics have been implemented by using JDeodorant [29] , and other tools are used to apply a popular approach in software measurement called feedback loop in extracting, evaluating of metrics values and and executing the obtained recommendations [34] through using McCabe IQ [24], and  Logiscope [26] . Also, MARFCAT tool [30] is utilized in order to detect code weaknesses of java files for both case studies. Finally, we present the obtained results and main points of this study.

The paper is organized as following. Section 1 gives a background on the two case studies and some software measurement metrics. Section 2 describes and interpret the result of using Logiscope and McCabe IQ on the two case studies. Section 3 presents the implementation of Chidamber and Kemere (C \& K) Metrics. Finally, Section 4 presents the results and conclusions.

\section{Background}
\subsection{OSS Case Studies}
\subsubsection{MARF}

Modular Audio Recognition Framework (MARF) is an open source framework. It implements number of supervised and unsupervised machine learning (ML) algorithms for the pattern recognition purpose [1]. Furthermore, MARF has some Natural Language Processing (NLP) along with network-related implementation. Furthermore, it is designed to be usable and easy configurable [2].  
 
The main objective of this platform is to provide the research community with a general machine learning and NLP framework. Another objective for MARF is to have the ability to extend its functionality. So, its users select from collection of ML and NLP algorithms, and compare between them [2].

MARF$'$s architecture is a pipeline that is written in Java. “Fig. 1” illustrates MARF$'$s architecture and how data stream through different stages. The incoming sample first loaded and stored into wav, mp3 or any type of files. The preprocessing stage then starts through normalization and be ready for feature extraction. By the end of extracting features, the data either considered as training data been stored or classified and the system lean from it. The strength of using this kind of architecture is the ability to add any number of algorithms between any two different stages [1, 2, 3].

 \begin{figure} [ht!]
  
  \centering
    \includegraphics[width=0.5\textwidth]{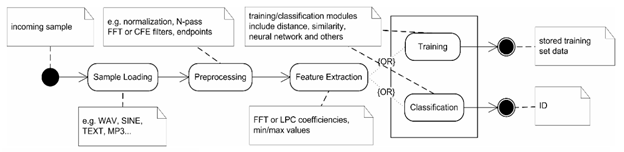}
\caption{MARF$'$s Pattern Recognition Pipeline [1, p. 474]}
\end{figure}

MARF has been utilized in number of applications. As an example of the extendibility feature of MARF in [1] numbers of plug-ins for sample loading and preprocessing have been implemented to identify writers. Another use of MARF is an attempt to figure out the most suitable algorithm combination for speech processing tasks [4]. In addition, MARF can run distributive on a network, but by that it may arise some of the security issues and in [5] suggested some solutions to most of the problems. First, confidentiality aspect where it concerns on correctness and accuracy of the data and the computation. To resolve this aspect is to use Java Data Security Framework (JDSF) confidentiality sub-framework. Second aspect is integrity where it concerns on getting a correct result when making a computation which can be achieved by using JDSF$'$s integrity sub-framework. Third aspect is authentication which is really important especially when it is used to identify a speaker or user which can resolved by using JDSF$'$s authentication sub-framework. The last aspect is availability of services, here most systems tend to create redundancy to achieve this aspect, but by that it may have a big risk to have a malicious attack. The proposed solution which is JDSF cannot resolve all security aspects, like the availability aspect, but it covers most of them [5]. Also, MARF and its NLP applications have been utilized to detect weaknesses and vulnerabilities in source code [6].

Finally, the main design of MARF aims to provide most of the best software qualities. First, the paper in [1] illustrates MARF$'$s modification and expandability quality through the implementation of extra components and integrate them to the system as plug-ins which means that it has the portability quality factor. While the usability factor provided through the variety of implemented algorithms which are easy to combine as needed. Also, MARF allows developers to change its components easily and this leads that MARF have the maintainability factor. In addition, it provides researchers with and accurate computation and processing and that means it has the functionality factor [2].


\subsubsection{GIPSY}

 The general intensional programming system (GIPSY) is a framework for the compilation and distributed demand-driven evaluation of context-aware declarative programs where the Intentional programming is defined as multidimensional context programming [12].
More than compilation, GIPSY can execute multiple programming languages, like Lucid, C++ and Java programming language [10]. GIPSY is a complex system that can bear redundant tiers to handle high volume access and provide high scalability to the system [8].

To understand the overall infrastructure, several quality criteria were stated prior designing GIPSY:
\begin{itemize}
\item  Adopting notation of language independence which allows to the execution of programs written in any language in the Lucid family;
\item  Adopting notation of scalability which allows to provide an architecture can work an efficient way and support distributed computing;
\item  Adopting notation of flexibility of execution architecture which allows to give and enhance an infrastructure;
\item  There are vagueness of run-time considerations through lack of run-time architectural and execution considerations can be candidly expressed into the source programs;
\item  Observation of execution infrastructure which makes for observable of execution parameters of all execution components , making experimentations to be made in order to reinforce the system and its execution for particular computational situations [7].
\end{itemize}

First GIPSY$'$s infrastructure will be demonstrated as how it meets these quality criteria by addressing the composition of GIPSY and its subcomponents. Then, an architecture overview will depict the most essential elements of the main components.

GIPSY is composed of three component or sub-systems (Figure 1).

 \begin{figure} [ht!]
  
  \centering
    \includegraphics[width=0.5\textwidth]{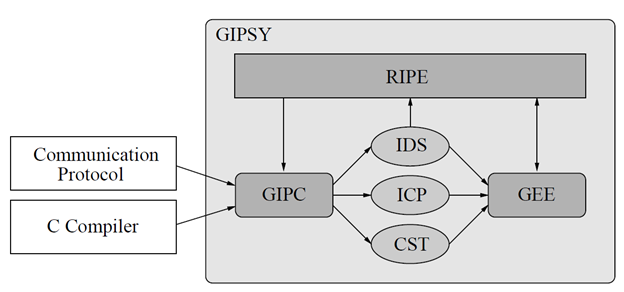}
\caption{The GIPSY Architecture [2, p. 148]}
\end{figure}

First component is the General Intentional Programming Language Compiler (GIPC). It is a two phase component that first translate the GIPSY program into intermediate Java. After that, the generated C code is compiled normally. In the second phase, Lucid is compiled into an intentional data dependency structure (IDS). Second component is the General Eduction Engine (GEE) which is demand-driven computation framework. This part deals with a procedure call that can be a locally or remotely, and it reduces the overhead related to procedure calls. Finally the Intentional Run-time Programming Environment (RIPE). In this module the dataflow diagram is generated. The design and implemented GIPSY architecture attempts to meet the following objectives: generality, adaptability, and efficiency [11].

Having languages with different data types would create an issue when they need to communicate and exchange variables. Declaring data type is another problem since they are implicit in Lucid and do not have to be declared in the syntactical level, unlike imperative languages. The primary goal is to allow Lucid to call and receive from methods in imperative Languages and the other way around, and also to allow programmers to be able to declare the data type in both intensional and imperative languages. A proposed solution is to have a bridge programming language between the intensional and imperative languages to declare data types [10]. 

The refactoring that was performed on GIPSY allowed to hide the specificities of each implementation by providing common interfaces ITransportAgent and IDemandDispatcher that each technology should implement, which increases the understandability of the system and moreover its extensibility to any other DMS technology. Also, this refactoring allows modularity and reusability since demand workers and generators source code has been extracted and is now common to both technologies. Other benefit of the actual refactoring is the ability of concurrent use of both technologies, which provide a suitable way for comparison [12]. 

Prior to the proposed solution in [9], GIPSY, was managed using a command-line interface. The project provides an GUI integrated tool create a GIPSY network, configure, and manage its components (GIPSY instances, tiers and nodes), dynamically visualize GIPSY nodes and tiers and inspect/change their properties at run-time, increase the usability of GIPSY run-time system as a whole and inspect the status and properties of GIPSY [9].

In the light of what has been mentioned above, the design elements criteria have been fulfilled as follows:
\begin{itemize}
\item The authors of [9] created a multi-tier architectural design which follows earlier solutions and enhances on the generator-worker architecture developed;
\item  Having a demand process where the demand propagation goes through tiers utilizing asynchronous messaging with message persistence , where demand stores could be queries and distributed on the distributed stores which can be solved via utilizing a peer-to-peer mechanism;
\item  a controller-based design, where each tier is administered via a controller/monitor. The efforts were made to design a new system with similar capacities, more flexibility, ability to deal with the fast evolution and diversity of the Lucid family of languages, and a language-independent run-time system for the execution of Lucid programs [7].
\end{itemize}

In conclusion, the previously mentioned design of GIPSY aimed to provide many quality attributes. Of these attributes we can mention: flexibility, and the ability to evolve in a fast manner (which means better maintainability), usability, extensibility, modularity, reusability understandability, generality, adaptability, efficiency, and scalability.

\subsection{Summary}

 The required measurements are presented in table \ref{tab:1} as follows:

 \begin{table}[htpb]
    
  \renewcommand{\arraystretch}{1.3}
 \caption{ MARF And GIPSY Measurements} 

    \begin{tabular}{|c|c|c|}

    \hline
    Measurements & MARF & GIPSY\\ \hline
   Files  &  201   &  601 \\ \hline

    Packages  & 46& 106       
 \\ \hline
 
  Classes  & 216& 665       
 \\ \hline

  Methods  & 2372& 6261       
 \\ \hline
  Line of Text  & 52833& 140247       
 \\ \hline

  Number of statements  & 7856& 56035       
 \\ \hline
  Comment Lines  & 13433 & 13606       
 \\ \hline
 
  Programming Languages  & IDL, Java, Perl, Unix Shell & Java      
 \\ \hline

    \end{tabular}

\label{tab:1}

 \end{table}

Table \ref{tab:1} , shows the measurements of both MARF and GIPSY projects. It is obvious that GIPSY project is larger than MARF since the number of files of GIPSY is three times more than MARF. Also, the number of classes and their corresponding method are following the same ratio. However, the number of statements in GIPSY is eight times greater than in MARF. Nevertheless, GIPSY has higher number of statements, MARF has been written in many programming language such as IDL, Java and Perl. Both GIPSY and MARF have nearlly the same number comment lines.

The methodology used to analyze both MARF and GIPSY projects source code was simple by using a tool called SonarQube [13]. SonarQube (previously known as Sonar [14]) is an open source tool for Inspection of code quality. It supports many languages such as Java, C/C++, PHP, JavaScript, Python, etc. [15].

We used windows 8.1 platform machine that has XAMPP v.1.8.3, which is an integrated server package of Apache, mySQL, PHP and Perl, [16] already installed. SonarQube has Sonar Server v.3.4.1 and Sonar Runner v2.0. To analyze both project, we first started Sonar Server which requires Apache HTTP Server to be running. We then created a sonar-project-properties file which contains the project source code files and the programming language used for each project. Consequently, we access the project folder from command line to run Sonar Runner.  Next, Sonar Runner read the project files and gathered the require measurement. As result, Sonar Server conducted the values and presented it in the browser.    

 \section{Metrics}
\subsection{MOOD Metrics}
In [17] Authors tried to evaluate all six object-oriented software metrics (MOOD) and investigate on their theoretic validity. Since these metrics could give us important information about software architecture, there is a need to know if the measures produce accurate and meaningful data to analysis.
Each of the measurement type possesses conditions that would allow to evaluate their related metrics : types direct and indirect, and the predicate that conditions are met would imply a theoretic validity.
MOOD consists of six metrics that allow assessing design properties such as encapsulation, inheritance, coupling, and polymorphism. 

First, encapsulation can be measured by two metrics Method Hiding Factor (MHF) and Attribute Hiding Factor (AHF).  MHF is defined as:

\begin{equation}
\frac{\sum_{i=1}^{TC}  \sum_{m=1}^{M_{d}(C_{i})} (1 - V(M_{mi}))}{\sum_{i=1}^{TC}  M_{d}(C_{i})}
\end{equation}

Where:
\begin{equation}
V(M_{mi})= \frac{ \sum_{j=1}^{TC}  is-Visible(M_{mi},C_{j}) } {TC-1}
\end{equation}

$M_{d}(C_{i})$ is the number of methods declared in a class, 
and TC is the total number of classes. $is-Visible(M_{mi},C_{j})=1$ When $i\neq j$, and $C_{j}$ can call $M_{mi}$. Otherwise it would get zero. The result will be the percentage of hidden methods in the system. AHF has the same MHF formal, but it uses attributes instead of methods.  In [17], they assumed that MHF and AHF meet three out of the four conditions that were proposed by Kitchenham for the direct measurements. 
The unsatisfied condition concerns the quality of information hiding design. 
Nevertheless, these two metrics are to measure the quantity of information hiding. Therefore, MHF and AHF do meet all conditions and so they are valid theoretically. In addition, they meet the indirect measurements conditions.

Second, inheritance can be measured in theory by two metrics listed as follow: Method Inheritance Factor (MIF) and Attribute Inheritance Factor (AIF) and they are defined as follow:

\begin{equation}
\frac{\sum_{i=1}^{TC}  M_{i}(C_{i})} {\sum_{i=1}^{TC}  M_{a}(C_{i})}
\end{equation}

And

\begin{equation*}
M_{i}(C_{i})= M_{d}(C_{i})+M_{i}(C_{i})
\end{equation*}

Where $M_{a}(C_{i})$ is the number of methods that can be called in association with $C_{i}$, $M_{d}(C_{i})$ is the number of methods declared in a class, and $M_{i}(C_{i})$ is the number of methods inherited and not overridden in $C_{i}$. For MIF, each class that inherits a method will be counted as one, and if not, it will be counted as zero. AIF has the same MIF formal, but it uses attributes instead of methods. MIF and AIF are only direct measurers, and therefore the writers in [17] compare it with the direct conditions. They found that MIF and AIF both are valid metrics.
 
Third,  Coupling Factor (CF) can hypothetically measure the coupling between classes in the system, but not coupling through inheritance.  It is defined as follows: 
\begin{equation}
\frac{\sum_{i=1}^{TC} \sum_{j=1}^{TC}  is-Client(C_{i},C_{j})} {TC^{2}-TC}
\end{equation}

$is-Client(C_{i},C_{j})$ is counted as one when there is a relation either by message passing or by semantic association links between the two classes, and $C_{c} \neq C_{s}$. $CF$ is direct and indirect measure. It can directly measure the coupling between classes, and indirectly measure complexity, lack of encapsulation, lack of reuse potential, lack of understandability, and lack of maintainability. The researchers came with a conclusion that CF as a direct measure is valid, but as an indirect measure, it is difficult to say it is valid, and therefore empirical evidence is needed.

Finally, Polymorphism Factor (PF) can potentially measure polymorphism potential in a system, and is defined as follow:

\begin{equation}
\frac{\sum_{i=1}^{TC} M_{o} (C_{i})} { \sum_{i=1}^{TC} \left [ M_{n}(C_{i}) * DC(C_{i}) \right ]}
\end{equation}

Where
\begin{equation*}
M_{d}(C_{i})= M_{n}(C_{i})+M_{o}(C_{i})
\end{equation*}

and $M_{n} (C_{i}) =$ the number of new methods,
$M_{o} (C_{i}) =$ the number of overriding methods,
$DC(C_{i})$ = the number of classes descending from $C_{i}$.

PF counts the number of inherited methods that are overridden in the system to measure polymorphism. Therefore, PF is an indirect measure. The researchers concluded that PF is not a valid metric, and therefore empirical evidence is needed.
 
In conclusion, MOOD metrics allow providing a high level view of the system architecture. Considered as being valid measures in the context of this theoretical, 
as it mentioned in [7] , using this metrics cannot be completely reliable unless complete empirical validations have been performed [17].


\subsection{QMOOD Metrics}
 Even if the need for quality has never been so important, the way to measure it is still not settled definitely.
With the raise of the object-oriented paradigm, new metrics have to be defined in order to satisfy different notions brought by the object-orientation such as encapsulation, inheritance and polymorphism.
Recent quality models as the one developed by Dromey addresses the limitations of previous models by decomposing the high-level quality attributes into substantial product quality properties, in a bottom-up fashion ensuring that the lower details are well specified.
Inspired from Dromey’s model, the methodology used in the development of the hierarchical quality model for object-oriented design (QMOOD) involves four steps. 
The first step aims of identifying design quality attributes. It consists of reworking the ISO 9126 attributes to adapt them for an oriented-object design such as  extendibility and reusability.
The purpose of the second step is the identification of the design properties, intrinsic to the oriented-object such as coupling and cohesion. Also, It has been noticed that the design properties can have positive or negative influence on the quality attributes as depicted by Table \ref{tab:2} , so they should be weighted proportionally in order to obtain all design properties at the same range.
The goal of the third step is the identification of the object-oriented design metrics that can be assessed by some object-oriented design properties as shown by Table \ref{tab:3} . Only metrics that can be applied at the design stage without requiring implementation has been retained such as direct class coupling metric (DCC) and cohesion among methods of class metric (CAM) as described by Table \ref{tab:4} .
The fourth step aims to identifying object-oriented design components which are essentially objects, classes and its structural and communication relationship between classes.

One of the strengths of the QMOOD model is the ability to be refined and adapted in accordance to the goals and objectives set, by changing its metrics for instance.
Finally, the present article exposes the procedures of validation that allowed to confirm the effectiveness of the QMOOD model of assessing reliably the overall quality of a software product.

 \begin{table}[htpb]
    
  \renewcommand{\arraystretch}{1.3}
 \caption{ Quality Attributes - Design Property Relationships}

    \begin{tabular}{|c|c|c|}

    \hline
      & Reusability & Extendibility\\ \hline
   Design size  &   \checkmark    &    \\ \hline

    Hierarchies  &  &         
 \\ \hline
 
  Abstraction  &  & \checkmark       
 \\ \hline

  Encapsulation  &  &         
 \\ \hline
  Coupling &  &         
 \\ \hline

 Cohesion & \checkmark &         
 \\ \hline
  Composition &   &         
 \\ \hline
 
  Inheritance  &  & \checkmark      
 \\ \hline

  Polymorphism  &  & \checkmark      
 \\ \hline

 Messaging  & \checkmark &       
 \\ \hline

 Complexity  & \checkmark &       
 \\ \hline

 \end{tabular}
  \label{tab:2}

 \end{table}

 \begin{table}[htpb]
    
  \renewcommand{\arraystretch}{1.3}
 \caption{ Design Metrics for Design Properties}

    \begin{tabular}{|c|c|} 

    \hline
   Design Property   & 
Derived Design Metric  
 \\ \hline
   Coupling  &    Direct Class Coupling (DCC)     \\ \hline

    Cohesion  & Cohesion among methods in Class (CAM)
   
 \\ \hline

 \end{tabular}
   \label{tab:3}
 \end{table}

 \begin{table}[htpb]
    
  \renewcommand{\arraystretch}{1.3}
 \caption{Design metrics Descriptions}

    \begin{tabular}{|c|p{1.5cm}|p{4cm}|} 

    \hline
   Metric   & Name & Description\\ \hline
   DCC  &   Direct class Coupling    &   The metric is a count of the different number of classes that a class is directly related to, including those related  by attribute declarations and message passing (parameters) in methods \\ \hline

    CAM  & Cohesion among methods of class &    This metric computes the relatedness among methods of a class based upon the parameter list of the methods. The metric is calculated using the summation of the intersection of parameters of  a method with the maximum independent set of all parameter types in the class. A metric value close to 1.0  is preferred (Range 0 to 1)     
 \\ \hline

 \end{tabular}
  \label{tab:4}
 \end{table}

In this paper, “An Empirical Study of the Relationship of Stability Metrics and the QMOOD Quality Models over Software Developed Using Highly Iterative or Agile Software Processes”, (2007), authors’ were worked to find out whether there were interesting relationships between the System Design Instability with entropy (SDIe) metric presented by Olague et al., the System Design Instability (SDI) metric developed by Li et al, and the Total Quality Index (TQI) described by Bansiya and Davis in six different highly iterative projects with multiple iterations [19], all of the projects were implemented in C++ using the object oriented paradigm. Five of these projects (A-E) were developed by senior students in undergraduate level of software engineering classes were taught by two different professors at two different universities, development process used was Extreme Programming “XP”, while the sixth project (F) was a highly iterative open source project, and its development process follows the 12 principles of agile methods [58]. After the authors’ has clarified the notation of Stability and its metrics as a measure of the ease of software evolution [19] and demonstrated some studies that were proposed metrics for stability as follows:

\subsubsection{SDI Metric} which was proposed the System Design Instability (SDI) metric by Alshayeb and Li. and measure of the evolution of object-oriented software from one iteration to another [58].it used the sum of three values: the percentage of classes whose names change from one iteration to the next, the percentage of class which were added and the percentage of classes which were deleted and how its value could be used to provide as indications regarding to the progress of the project and how its value could be helpful to modifying the plan of the project during development process [19].

\subsubsection{SDIe Metric} which was proposed as a revision to the SDI metric by Olague et al. It is computed by using the number classes added, deleted, changed and unchanged from the previous iteration to another and this metric was examined by using theoretical validation [19].

\subsubsection{Bansiya Quality Model} which was proposed by Bansiya and Davis to define a hierarchical model (known as QMOOD) for object oriented design quality assessment which used quality factors from object oriented designs [19]. It was based on selected set of quality factors of the ISO 9126. They have excluded some factors such as reliability, usability, efficiency, portability and maintainability which were not related to design quality and could be considered them related to implementation. Thus, they have included only functionality and have added the quality factors of reusability, flexibility, extendibility, effectiveness, and understandability because of their relation to object oriented design [58]. These six quality factors are not directly measurable, but Bansiya and Davis have identified eleven design properties as described in table \ref{tab:5}  which could be directly measured and used to derive the six quality factors [19].

 \begin{table}[htpb]
    
  \renewcommand{\arraystretch}{1.3}
 \caption{Design Properties.}

 \begin{tabular}{|c|p{4cm}|} 

    \hline
  Design property  & Property Description \\ \hline
  Design Size (DSC) &  A measure of number of classes used in the design [20].
   
 \\ \hline
  Hierarchies (NOH) &Number of class hierarchies in the design [20] and these classes in a design which may have one or more descendants exhibit. 
 \\ \hline

 Abstraction (ANA) & A measure of generalization- specialization aspect of design. This metric is equivalent to DIT (Depth of Inheritance Tree) from Chidamber and Kemerer [20].
 \\ \hline

 Encapsulation (DAM) & A measure can be calculated by computing the ratio of the number of private (protected) attributes to the total number of attributes declared in the class [20].
 \\ \hline

Coupling (DCC) & A measure can be calculated by count of the different number of classes that are directly related to other classes (by attribute declarations and parameters) [20].
 \\ \hline
 Cohesion (CAM) & “Accesses the relatedness of methods and attributes in a class. Strong overlap in method parameters and attribute types is an indication of strong cohesion” [19].

 \\ \hline

Composition (MOA) & A measure can be calculated by computing of the part-whole relationships realized by attributes [20].
\\ \hline

Inheritance (MFA) & A measure can be calculated by computing the proportion of the number of methods inherited and exclude overriding methods by a class over the total number of methods in the class (inherited + defined) [20].
\\ \hline

Polymorphism (NOP) & A measure can be calculated by computing of the number of abstract methods defined in the class [20].
\\ \hline
Messaging (CIS) & A measure can be calculated by computing the average of number of public methods that are available as services to other classes [59].
\\ \hline

Complexity (NOM) & A measure can be calculated by computing the average of all the methods defined in a class. [20].
\\ \hline

 \end{tabular}
  \label{tab:5}

 \end{table}

Regarding eleven design properties above, some of them had a positive relationship on a given quality factor, while the rests had a negative relationship [19]. The quality factors were computed via the direct measurements that were taken from the high level abstraction of UML paradigm, class diagram and Table \ref{tab:6} lists the formula relating the contribution of the design properties on the quality factors.

 \begin{table}[htpb]
    
  \renewcommand{\arraystretch}{1.3}
 \caption{QMOOD Quality Factors and Design Properties Relationships.}

 \begin{tabular}{|c|p{4.5cm}|} 

    \hline
  Quality Factor  & Relationship \\ \hline

  Reusability &  - 0.25 * Coupling + 0.25*Cohesion +0.5*Messaging +0.5*Design Size
   
 \\ \hline
  Flexibility & 0.25*Encapsulation -0.25*Coupling +0.5*Composition +0.5*Polymorphism
 \\ \hline

Understandability & -0.33 * Abstraction + 0.33 * Encapsulation - 0.33*Coupling + 0.33*Cohesion -0.33*Polymorphism - 0.33*Complexity -0.33* Design Size
 \\ \hline

  Functionality &  0.12*Cohesion + 0.22*Polymorphism + 0.22*Messaging + 0.22*Design Size + 0.22*Hierarchies
 \\ \hline

 Extendibility &  0.5*Abstraction -  0.5*Coupling + 0.5*Inheritance + 0.5* Polymorphism
 \\ \hline
  Effectiveness & 0.2*Abstraction + 0.2*Encapsulation + 0.2*Composition + 0.2*Inheritance +0.2*Polymorphism

 \\ \hline
 
 \end{tabular}
  \label{tab:6}
 \end{table}

The summation of the quality factors was computed to reach the overall QMOOD total quality index (TQI). 
Therefore, $TQI = Reusability + Flexibility + Understandability + Functionality + Extendibility + Effectiveness$ [19].

Finally, from the results of the statistical analysis which were obtained from this empirical study, there is a positive relationship between the both of the stability metrics and TQI value. There is obviously a stronger correlation between the TQI value and the SDI metric than between the TQI value and SDIe [19].

\subsection{Coupling Metric}

The paper [21] propose a framework to choose the suitable coupling metrics based on the measurement goal. A survey and study have been conducted to review of the coupling metrics exist in lectures. The coupling metrics mentioned in the paper and used in the framework are:

\subsubsection*{{\bf Coupling between object (CBO)}} it measures the number of other classes that are coupled to a specific class, and it does not count the inheritance relation. There are two version of this metrics (CBO and CBO’) as follows.

\begin{equation}
CBO(c) = |\left \{ d \in C - \left \{ c \right \}|uses(c,d)\vee uses(d,c) \right \}|
\end{equation}

\begin{multline}
$$CBO'(c) = | \{ d  \in  C - (\{c\} \cup Ancestors (C))| \\ uses(c,d) \vee uses(d,c)\}|$$
\end{multline}

From its definition, class c coupled to class d if it uses class d or being used by class d. Thus, CBO counts the used of both methods and attributes. The most important characteristic of CBO metrics is that it does not distinguish between import and export coupling.

\subsubsection*{{\bf Response for class (RFC)}} the coupling measures based on method-method interaction. That is, for a class it count all methods invoked by all methods define in this class. It is define as $RFC = |RS|$ where $RS$ is the response set for the class, and the class response set can be expressed as

\begin{equation}
RS = {M} \cup_{all\, i} \left \{ R_{i} \right \}
\end{equation}

where ${R}$ is the set of methods called by method I, and ${M}$ is the set of all methods in the class. However, in [21] a new version of RFC is defined as following:

\begin{equation}
RFC_{\alpha }\left (c \right )=|\cup _{i=0}^{\alpha }R_{i}\left (c \right )|, $$For$$  \alpha = 1,2,3,...,
\end{equation}

It takes into account the levels of nested methods invocation.

\subsubsection*{{\bf Message passing coupling (MPC)}} it measures coupling based on method invocations. It counts the number of send statement define in a class. It is the number of static invocations of methods not the ones implemented in as shown in the following equation.  
\begin{equation}
MPC\left ( c \right )= \sum _{m\in M_{I}\left ( c \right )}  \, \sum _{m'\in SIM\left ( m \right )-M_{I}\left ( c \right )} NSI\left ( m,m{}' \right )
\end{equation}

\subsubsection*{{\bf Data abstraction coupling (DAC)}} it measure the coupling by counting numbers of abstract data types defined in a class. In this case, abstract data type is basically a class define in the system. The metrics definition is shown in the following equation:
\begin{equation}
DAC\left ( c \right )= |\left \{ a| a \in A_{I}\left ( c \right )\wedge T\left ( a \right )\in C \right \}|
\end{equation}

\begin{equation}
DAC{}'\left (c\right )= |\left \{ T\left (a\right ) |a\in A_{I}\left ( c \right ) \wedge T\left ( a \right ) \in C\right \}|
\end{equation}

$ DAC\left ( c \right )$  counts the number of abstract data type used in class c, while $ DAC{}'\left ( c \right )$  gives the number of other class using class c. In fact, this metrics take into account aggregation.
\subsubsection*{{\bf Ce and Ca (efferent and afferent coupling)}} this method categories classes into groups that achieve some common goal. The original definition of Ce and Ca is:
Ca: the number of classes outside this category that depend upon classes within this category.
Ce: the number of classes inside this category that depend upon classes outside this category.
 
\subsubsection*{{\bf Coupling factor (COF)}} it measures the coupling between the classes of a system. Suppose the total number of classes in a system is TC, and isclient(cc,cs) is function that return 1, if class cc is not a descendent of class cs, and cc ≠ cs and cc references a method or attribute of class cs. Otherwise, inclient(cc,cs) return 0. The original definition of COF is:
\begin{equation}
COF\left ( S \right )= \frac{\sum ^{TC}_{i=1}\sum ^{TC}_{j=1}isclient\left ( c_{i},c_{j} \right )}{TC^{2} - TC - \left [ 2\sum ^{TC} _{i=1} |Descendents\left ( c_{i} \right )|\right ]}
\end{equation}

Again, COF does not distinguish between method invocations and attribute reference. 
Information-flow-based coupling (ICP): for a method M in a class c, it counts the number of polymorphistically invoked methods of other classes, weighted by the number of the invoked method.

\begin{multline}
$$ICP^{c}\left ( m \right )= \sum _{m{}'\in PIM\left ( m \right )-\left ( M_{NEW}\left ( c \right )\cup M_{OVR}\left ( c \right ) \right )} \\ \left ( 1+|Par\left ( m{}' \right )| \right )\cdot  NPI\left ( m,m{}' \right )$$
\end{multline}

This metrics can be applied to classes and a subsystem as follow:
\begin{equation}
ICP\left ( c \right )=\sum _{m\in M_{I}\left ( c \right )}ICP^{c}\left ( m \right )
\end{equation}

\begin{equation}
ICP\left ( SS \right )= \sum _{c\in SS}ICP\left ( c \right )
\end{equation}

A Unified Framework for Coupling Measurement: the main goal of this framework is to provide a guideline for comparing and choosing coupling measures. It define six criteria that affect the measurement goal. The framework’s criteria are: 
1) type of connection such as attribute, method call or both. 
2) the locus of impact if it is import or export coupling.
3) the granularity of the measure which includes specifying the domain and components are to be measured. 
4) stability of server which means if the class is subject to development of change in the project.
5) direct or indirect coupling, which means to count the direct or indirect connection between classes.
6) inheritance based measure and non-inheritance base that mean to take into account the coupling based on inheritance.


\subsection{Cohesion Metric}

In their paper “Measurement of Cohesion and Coupling in OO Analysis Model Based on Crosscutting Concerns,” O.Ormandjieva, M. Kassab, C. Constantinides (2005) introduce measurements for controlling the coupling and the cohesion of the Object-Oriented analysis based on the notion of crosscutting concerns (or aspects). The measurement control mechanism helps to identify early the implication of crosscutting in the system. There are shared sub-functionality between two or more use-cases, known as crosscutting, which leads to lower cohesion in the use-case model.  To measure the cohesion of the system, the authors proposed a new model which is based on use-case scenarios. To be able to measure the Cohesion Level for a given user-case, the paper presents a new formulas as follows:

\begin{equation}
CL\char`_UC= \frac{|Q|} {|P|}
\end{equation}

Where Q is the set of the similar pairs of scenarios belonging to one use-case, and P is the set of all pairs of scenarios belonging to the same use-case.
The $CL-UC$ value threshold is between 0 and 1, where 1 shows the highest level of cohesion.
 
The Cohesion Level in the Use-Case Model measure is defined on the set of all scenarios belonging to all use-cases in the use-case model:
\begin{equation}
CL\char`_UCM= 1-\frac{|QM|} {|PM|}
\end{equation}

Where $QM$ is the set of the pairs of similar scenarios and $PM$ is the set of all pairs of scenarios.
The $CL\char`_UCM$ value threshold is between 0 and 1, where 1 shows the highest level of cohesion.

For coupling design property, they have adopted the MOOD’s Coupling Factor measure to quantify the level of coupling in the domain model.
\begin{equation}
FC=\frac{ \sum_{i=1}^{TC}  \sum_{j=1}^{TC} \texttt{is\char`_client}(C_{i},C_{j}) }{TC^{2} - TC}
\end{equation}

Where

\begin{equation}
 is\char`_client(C_{i},C_{j}) = \begin{cases}
 & \text{ 1 iff } C_{c} \Rightarrow C_{s} \wedge C_{c} \neq C_{s} \\
 & \text{ 0  } otherwise
\end{cases}
\end{equation}

$C_{c} \Rightarrow C_{s}$ represents the relationship between a client class $C_{c}$ and a supplier class $C_{s}$

The CF value threshold is between 0 and 1, where 1 shows the highest level of cohesion.
 
Both cohesion formulas were validated theoretically against Formal Properties of Cohesion. Coupling also was verified against Formal Properties of Coupling [22].

In the paper “A Unified Framework for Cohesion Measurement in Object-Oriented Systems” the authors discuss the existing cohesion measurements in Object-Oriented Systems and the difficulty to determine how such measures relate to one another, for which application they can be used, and how to select or define these measures. A unified framework is introduced and all existing measures are then classified according to this framework [23]. The authors first present the already existing measurement approaches:

\begin{itemize}
\item Chidamber and Kemerer: LCOM1 (Lack of Cohesion in Methods), LCOM2
\item Hitz and Montazeri: LCOM3, LCOM4, Co (Connectivity ‘C’ in original definition)
\item Bieman and Kang: TCC (tight class cohesion), LCC (loose class cohesion)
\item Henderson-Sellers: LCOM5
\item Lee et al.: ICH (Information-flow-based Cohesion)
\item Briand et al.: RCI (Ratio of cohesive interactions), NRI (Neutral RCI), PRCI (Pessimistic RCI), ORCI (Optimistic RCI)

\end{itemize}

Here is the definition of some of these measures:
 
\subsubsection{LCOM2} is the degree of similarity of methods by instance variable or attributes such that, having P (the set of method pairs that do not share attributes) and Q (the set of method pairs that do share at least one attribute).
\begin{equation} 
LCOM2\left ( c \right )=\left \{_{0,\ otherwise}^{|P| - |Q|,\ if |P| > |Q|}  \right.
 \end{equation}

\subsubsection{LCOM5}

\begin{multline}
$$LCOM5\left ( c \right )= \\ \frac{|M_{I}\left ( c \right )| - \frac{1}{|A_{I}\left ( c \right )|}  \sum_{a\in A_{I}\left ( c \right )}|\left \{ m|m \in  M_{I}\left ( c \right ) \wedge  a \in AR\left ( m \right ) \right \}|}{|M_{I}\left ( c \right )| - 1}$$
\end{multline}

Where:
\begin{itemize}
\item  AR(m) the set of attributes referenced by method ‘m’,
\item MI (c) the set of method implemented in class ‘c’, 
\item AI (c) is the set of attributes implemented in class ‘c’,
\item m is a method and a is an attribute of class ‘c’
\end{itemize}
 
\subsubsection{TCC} is defined as the percentage of pairs of public methods of the class with common attribute usage

 \begin{equation}
\resizebox{1\hsize}{!}{$ TCC(c)= 2 \frac{|\{\{ m1,m2 \}|m1,m2  \in M_{I}(c)\cap M_{pub}(c)\wedge m1  \neq m2 \wedge cau(m1,m2) \}|}{|M_{I}(c)\cap M_{pub}(c)|(|M_{I}(c) \cap M_{pub}(c)|-1)}$}
\end{equation}

Where: 
\begin{itemize}
\item $M_{pub}(c)$ is the set of public methods of ‘c’ and
\item $cau(m_{1},m_{2})$ (common attribute usage) which is true, if m1, m2 M1(c) directly or indirectly use an attribute of class c in common.

LCOM4: let $G_{c} = (V_{c}, E_{c})$ be an undirected graph with vertices $V_{c} = M_{I}(c)$  edges $E_{c} = \{ \{ m_{1}, m_{2} \} | (m_{1}, m_{2} \in V_{c})$ and $(AR(m_{1}) \cap AR(m_{2}) \cap AI(c) \neq  \phi$ OR $m_{1} \in SIM(m_{2})$ Or $m_{2} \in SIM(m_{1})) \}$  
LCOM4(c) is the number of connected components of $G_{c}$

 Where:
\item SIM(m) the set of statically invoked methods of m

\end{itemize}

Then, the defined framework should ensure the following:
\begin{itemize}
\item  measure definitions are based on explicit decisions and well understood properties,
\item  all relevant alternatives have been considered for each decision made,
\item dimensions of cohesion for which there are few or no measures defined are highlighted.
 \end{itemize}

The framework consists of five criteria, each criterion determining one basic aspect of the
resulting measure:
\begin{itemize}
\item  The type of connection, i.e., what makes a class cohesive.
 \item     Domain of the measure.
 \item      Direct or indirect connections.
 \item    Inheritance: how to assign attributes and methods to classes, how to account for polymorphism.
 \item     How to account for access methods and constructors.
 \end{itemize}

These criteria helps deriving and building the framework that shows what cohesion measures should be used. As an example from the proposed framework in [23], the first three criteria and their corresponding tables are presented to show options if any, which will be strongly influenced by the stated measurement goal:

 \begin{itemize}
\item  Type of connection: It is the mechanism that makes a class cohesive. Table \ref{tab:99} shows the types of connections between the elements of a class (attributes, methods, or data declarations)
 \end{itemize}

 \begin{table*}[!t]
    
  \renewcommand{\arraystretch}{1.3}
 \caption{Types of connection [23].}

  \begin{center}
 \begin{tabular}{|c|c|c|c|c|}
\hline
Element 1 & Element 2 & Description &  \vtop{ \hbox{ \strut Design} \hbox{ \strut phase}} & Measures \\ \hline
method m of class c & attribute a of class c & m references a & HLD & LCOM5 \\ \hline
method m of class c & method m$'$ of class c & m invokes m$'$ & HLD & ICH, LCOM4, Co \\ \hline

method m of class c & \vtop{ \hbox{ \strut method m' of class c,} \hbox{ \strut  m $ \neq $ m' }}& \vtop{ \hbox{ \strut m and m' directly reference an attribute a } \hbox{\strut  of class c in common ("similar methods") }}& HLD &  \vtop{ \hbox{ \strut  LCOM1, LCOM3,}  \hbox{ \strut   LCOM4, Co, LCOM2 }} \\ \hline

method m of class c & \vtop{ \hbox{ \strut method m' of class c,}  \hbox{ \strut  m $ \neq $ m' }}& \vtop{ \hbox{ \strut m and m' indirectly reference an attribute a } \hbox{ \strut of class c in common ("connected methods") }}& HLD & TCC, LCC \\ \hline
data declaration in class c & \vtop{ \hbox{ \strut data declaration} \hbox{ \strut  in class c}} & data-data interaction & HLD & \vtop{ \hbox{ \strut RCI, NRCI,}  \hbox{ \strut  ORCI, PRCI}} \\ \hline
method m of class c &\vtop{ \hbox{ \strut data declaration} \hbox{ \strut  in class c}} & data-method interaction & HLD & \vtop{ \hbox{ \strut RCI, NRCI,} \hbox{ \strut  RCI, PRCI}} \\ \hline
 
\end{tabular}
 \end{center}

  \label{tab:99}

\end{table*}


\begin{itemize}
\item  Domain of the Measure: Domain of measure shows the objects that will be measured (methods, classes, etc…). Table \ref{tab:98} shows these possible domains.
\end{itemize}

\begin{table}[h]
 \caption{Domain of the measure [23].}
\begin{tabular}{|c|c|}
\hline
Domain     	& Measures                                                                                                          	\\ \hline
Attribute  	& $-$                                                                                                              	\\ \hline
Method     	& ICH                                                                                                               	\\ \hline
Class      	& \begin{tabular}[c]{@{}c@{}}LCOM1, LCOM2, LCOM3, Co, LCOM4, LCOM5,\\ TCC, LCC, ICH, RCI, NRCI, PRCI, ORCI\end{tabular} \\ \hline
Set of Classes & ICH                                                                                                               	\\ \hline
System     	& $-$                                                                                                              	\\ \hline
\end{tabular}
  \label{tab:98}
\end{table}

  

\begin{itemize}
\item   Direct or Indirect Connections: Table \ref{tab:97} shows which measures take only direct connection (connection type \#3 in Table \ref{tab:99} ) and which takes indirection connections (type \#4 in Table \ref{tab:99} ) into consideration.
\end{itemize}

\begin{table}[h]
\caption{Direct and Indirect Connection [23].}
\begin{tabular}{|c|c|}
\hline
Type 	& Measures                             	\\ \hline
Direct   & LCOM1, Co, LCOM2, LCOM5, TCC, ICH    	\\ \hline
Indirect & LCOM3, LCOM4, LCC, RCI, NRCI, ORCI, PRCI \\ \hline
\end{tabular}
 \label{tab:97}
\end{table}

  

\subsection{Summary}
The study of MARF and GIPSY shows these software systems have many qualities attributes MARF has several quality attributes such as expandability, portability, usability, maintainability, and functionality, while GIPSY has many quality attributes such as flexibility, usability, modularity, reusability, understandability, generality, adaptability, efficiency, and scalability. With the growing size of MARF and GIPSY software systems and that will lead to increase the complexity of these software systems. Thus, having significant measures can be computed without any difficulty is needed. These measures may help us to assess the quality attributes of both systems. However, this is still considered a very difficult and challenging task. 

In this section, a set of metrics is proposed that will be implemented in a future work, and then try to discover the relationship between these metrics and some of important quality attributes of these software such as maintainability, or some of the quality attributes mentioned above. Therefore, we have prioritized these metrics based on importance in term of assessing the quality of object-oriented of software system as following: 
MOOD metrics are well known metrics and used to measure some characteristics of the Object-Oriented programs that cover most of aspects of maintainability since object-oriented metrics address the several issues like coupling, inheritance, encapsulation and polymorphism etc. MOOD metrics can be listed as follows:

\subsubsection{Coupling related Metric}
Coupling is a measure of interdependence between the classes of a system. Two classes are coupled if methods of one use methods and/or instance variables of the other. The Coupling metric is related to many quality attributes such as usability, maintainability, reusability, testability, understandability etc are used to enhance the quality of software system. Coupling Factor CF metric will be part of the selected metrics in the future milestones.

\subsubsection{Cohesion related Metrics}
Cohesion refers to how closely the operations in a class are related to each other. The Cohesion metric is related to some quality attributes such as maintainability, reusability, testability, understandability, etc are used to assess the quality of software system. The Chidamber and Kemerer’s LCOM (Lack of Cohesion in Methods) metric is part of the future milestone.

\subsubsection{Encapsulation related Metrics}
Encapsulation metric measures how attributes or/methods are encapsulated in the classes of a system. It is related to some quality attributes such as maintainability, reusability, testability, understandability etc are used to assess the quality of software system. The Method Hiding Factor (MHF) and Attribute Hiding Factor (AHF) metrics will be part of the future milestones.

\subsubsection{Polymorphism related Metric}
Polymorphism metric measures the degree of method overriding in the class inheritance trees. It is related to some quality attributes such as maintainability, reusability, testability, understandability etc are used to assess the quality of software system. The Polymorphism Factor (PF) metric is part of the future milestone. 

\subsubsection{Inheritance related Metrics}

Inheritance metric measures how attributes or/methods are Inherited in the classes of a system. It is related to some quality attributes such as maintainability, reusability, testability and understandability that are used to assess the quality of software system. Method Inheritance Factor (MIF) and Attribute Inheritance Factor (AIF) metrics are among the selected metrics for the future milestones. 

The previous milestone related a refactoring work on both system could increase the maintainability factor and the extensibility of their components. It would be interesting to assess how well this refactoring will increase the maintainability attribute and reduced the coupling design property since coupling has a negative effect on maintainability. Then, it is straightforward that the coupling related metric (Coupling factor metric) would be prioritized over the other metrics.  Since a good design is not only related to a low coupling but also to a high cohesion, it is then essential that the cohesion related metrics would be ranked in a second priority (LCOM). An overview of the remaining design properties shows that the inheritance is the higher level property and is most likely to have an impact on the overall quality of the systems. Then, it is evident that the inheritance related metrics should be ranked in the third position in priority (Method Inheritance Factor (MIF) and Attribute Inheritance Factor (AIF) metrics). The encapsulation related metric have the fourth position in ranking in terms of influence on the general quality of the software(The Method Hiding Factor (MHF) and Attribute Hiding Factor (AHF)) . Finally, the polymorphism related metric has the last ranking due to it small effect on the overall quality of the software.

\section{Methodology}
\section*{Metrics with Tools: McCabe and Logiscope}
\subsection{Logiscope:}
The maintainability factor is “the capability of the software product to be modified.  Modifications may include corrections, improvements or adaptation of the software to changes in environment, and in requirements and functional specifications” [ISO/IEC 9126-1:2001]. The formula to compute the factor is mainly the sum of its criteria for all levels:

$Maintainability = Analyzability + Changeability +Stability + Testability$

From the maintainability equation above, LogisCope computes the maintainability factor based on four criteria which are listed below. These criterion’s definition and formula taken from the LOGISCOPE report documentation.
 
 {\it {\bf Analyzability:}} LOGISCOPE report documentation defines analyzability as “The capability of the software product to be diagnosed for deficiencies or causes of failures in the software, or for the parts to be modified to be identified” [ISO/IEC 9126-1:2001]. The formula to compute the criteria is:

$ANALYZABILITYc = cl\_wmc + cl\_comf + in\_bases + cu\_cdused$

{\it{\bf Changeability:}} LOGISCOPE report documentation defines changeability as “The capability of the software product to enable a specified modification to be implemented” [ISO/IEC 9126-1:2001]. The formula to compute the criteria is:

\begin{equation*}
CHANGEABILITYc = cl\_stat + cl\_func + cl\_data
\end{equation*}

{\it {\bf Stability:}} LOGISCOPE report documentation defines stability as “The capability of the software product to avoid unexpected effects from modifications of the software” [ISO/IEC 9126-1:2001]. The formula to compute the criteria is:

$STABILITYc = cl\_data\_publ + cu\_cdusers + in\_noc + cl\_func\_publ$

{\it {\bf Testability:}} LOGISCOPE report documentation defines testability as “The capability of the software product to enable modified software to be validated” [ISO/IEC 9126-1:2001]. The formula to compute the criteria is:

\begin{equation*}
TESTABILITYc = cl\_wmc + cl\_func + cu\_cdused
\end{equation*}

 \begin{table}[htpb]
    
  \renewcommand{\arraystretch}{1.3}
 \caption{Operands Description}

 \begin{tabular}{|c|p{4cm}|p{2cm}|} 
  \hline
Operand   & Definition                                                                                                                                                                                                                                                                       & Measured Attribute            	 \\ \hline
cl\_wmc    	& Sum of the static complexities of the class methods. Static complexity is represented by the cyclomatic number of the functions.                                                                                                                                                     	& Cohesion                       \\ \hline
cl\_comf   	& cl\_comf = (cl\_comm) / (cl\_line MAX 1)Ratio between the number of lines of comments in the module and the total number of lines:cl\_comf = cl\_comm / cl\_linewhere: cl\_comm is the number of lines of comments in the package, cl\_line is the total number of lines in the package. & Size . Reuse                   \\ \hline
in\_bases  	& Number of classes from which the class inherits directly or not. If multiple inheritance is not used, the value of in\_bases is equal to the value of in\_depth.                                                                                                                     	& Inheritance                    \\ \hline
cu\_cdused 	& The number of directly classes used by this class.                                                                                                                                                                                                                                   	& Coupling .\null  Cohesion            \\ \hline
cl\_stat   	& Number of executable statements in all methods and initialization code of a class.                                                                                                                                                                                                   	& Size                          	 \\ \hline
cl\_func   	& The number of methods declared in the class.                                                                                                                                                                                                                                         	& Size .  \null Cohesion               \\ \hline
cl\_data   	& The number of attributes declared in the class.                                                                                                                                                                                                                                      	& Size. \null Cohesion                 \\ \hline
cl\_data\_publ & The number of attributes declared in the class’ public section.                                                                                                                                                                                                                      	& Coupling. \null Cohesion. \null Encapsulation  \\ \hline
cu\_cdusers	& The number of classes that directly use this class.                                                                                                                                                                                                                                  	& Coupling. \null  Cohesion            	 \\ \hline
in\_noc    	& The number of classes which directly inherit the class the class.                                                                                                                                                                                                                    	& Inheritance                 \\ \hline
cl\_func\_publ & The number of methods declared in the class’ public section.                                                                                                                                                                                                                         	& Coupling. Cohesion. Polymorphism
\\ \hline

\end{tabular}
  \label{tab:7}

\end{table}

\subsubsection{Theoretical Validation}
We used a theoretical validation approach to prove each of tracking and consistency at different time and the direction of change should be in the same for both. Suppose we have java Project P that contains two Versions ver1, ver2 and we have some Metrics M for each one of four criteria level that are made up maintainability factor. Hence, these criteria and metrics are presented as in Table \ref{tab:96}:

\begin{table}[h]
\caption{Maintainability$'$s Criteria and Metrics}
\begin{tabular}{|p{1.6cm}|p{1.5cm}|p{4cm}|}
\hline
Factor                       	& Criteria  	& Metrics                                                         	\\ \hline
\multirow{4}{*}{Maintainability} & Analyzability & M\_cl\_wmc + M\_cl\_comf + M\_in\_bases + M\_cu\_cdused         	\\ \cline{2-3}
                             	& Changeability & M\_cl\_stat + M\_cl\_func + M\_cl\_data                         	\\ \cline{2-3}
                             	& Stability 	& M\_cl\_data\_publ + M\_cu\_cdusers + M\_in\_noc + M\_cl\_func\_publ \\ \cline{2-3}
                             	& Testability   & M\_cl\_wmc + M\_cl\_func + M\_cu\_cdused                        	\\ \hline
\end{tabular}
  \label{tab:96}
\end{table}


The purpose of using each metric from the table \ref{tab:7} above is described in Table \ref{tab:95}:

 \begin{table}[htpb]
    
  \renewcommand{\arraystretch}{1.3}
\caption{Metrics$'$ Definition}

 \begin{tabular}{|p{2cm}|p{4cm}|}  
\hline  	 
Metrics & Definition    \\ \hline
M\_cl\_comf               	& ratio between the number of lines of comments in the module and the total number of lines: cl\_comf = cl\_comm / cl\_line \\ \hline
M\_cl\_data               	& The number of attributes declared in the class.                                                                                                                 	\\ \hline
M\_cl\_data\_publ         	& The number of attributes declared in the class' public section.                                                                                                 	\\ \hline
M\_cl\_func               	& The number of methods declared in the class.                                                                                                                    	\\ \hline
M\_cl\_func\_publ         	& The number of methods declared in the class' public section.                                                                                                    	\\ \hline
M\_cl\_stat               	& Number of executable statements in all methods and initialization code of a class.                                                                              	\\ \hline
M\_cl\_wmc                	& Sum of the static complexities of the class methods.                                                                                                            	\\ \hline
M\_cu\_cdused             	& The number of directly classes used by this class.                                                                                                              	\\ \hline
M\_cu\_cdusers            	& The number of classes that directly use this class.                                                                                                             	\\ \hline
M\_in\_bases              	& Number of classes from which the class inherits directly or not.                                                                                                	\\ \hline
M\_in\_noc                	& The number of classes which directly inherit the class.                                                                                                         	\\ \hline
\end{tabular}
  \label{tab:95}
\end{table}


While we use Logiscope tool, we find that the values for these Metris within the range of acceptable values are organized as in Table \ref{tab:94} and Table \ref{tab:93}.

 \begin{table}[htpb]
    
  \renewcommand{\arraystretch}{1.3}
\caption{Version 1 Metrics$'$ Values}

\begin{tabular}{|l|l|l|l|l|}

\hline
\multicolumn{1}{|c|}{\multirow{2}{*}{Version}} & \multicolumn{1}{c|}{\multirow{2}{*}{Metric}} & \multirow{2}{*}{Value} & \multicolumn{2}{l|}{Range of acceptable Value} \\ \cline{4-5}
\multicolumn{1}{|c|}{}                     	& \multicolumn{1}{c|}{}                    	&                    	& Min                	& Max               	\\ \hline
\multirow{11}{*}{Ver1}                     	& M\_cl\_comf1                             	& 89                 	& 0                  	& 100               	\\ \cline{2-5}
                                           	& M\_cl\_data1                             	& 21                 	& 0                  	& 25                	\\ \cline{2-5}
                                           	& M\_cl\_data\_publ1                       	& 6                  	& 0                  	& 7                 	\\ \cline{2-5}
                                           	& M\_cl\_func1                             	& 6                  	& 0                  	& 9                 	\\ \cline{2-5}
                                           	& M\_cl\_func\_publ1                       	& 14                 	& 0                  	& 20                	\\ \cline{2-5}
                                           	& M\_cl\_stat1                             	& 5                  	& 0                  	& 7                 	\\ \cline{2-5}
                                           	& M\_cl\_wmc1                              	& 9                  	& 0                  	& 11                	\\ \cline{2-5}
                                           	& M\_cu\_cdused1                           	& 5                  	& 0                  	& 6                 	\\ \cline{2-5}
                                           	& M\_cu\_cdusers1                          	& 2                  	& 0                  	& 3                 	\\ \cline{2-5}
                                           	& M\_in\_bases1                            	& 3                  	& 0                  	& 4                 	\\ \cline{2-5}
                                           	& M\_in\_noc1                              	& 4                  	& 0                  	& 5                 	\\ \hline
\multicolumn{5}{|l|}{\vtop{ \hbox{ \strut Analyzability = M\_cl\_wmc + M\_cl\_comf + M\_in\_bases} \hbox{ \strut  + M\_cu\_cdused = 9+89+3+5=106}}}                                                        	\\ \hline
\multicolumn{5}{|l|}{Changeability = M\_cl\_stat + M\_cl\_func + M\_cl\_data =21}                                                                                   	\\ \hline
\multicolumn{5}{|l|}{\vtop{\hbox{ \strut Stability = M\_cl\_data\_publ + M\_cu\_cdusers + M\_in\_noc } \hbox{ \strut + M\_cl\_func\_publ =26}}}                                                           	\\ \hline
\multicolumn{5}{|l|}{Testability = M\_cl\_wmc + M\_cl\_func + M\_cu\_cdused = 20}                                                                                   	\\ \hline
\multicolumn{5}{|l|}{\vtop{\hbox{ \strut Maintainability = Analyzability + Changeability} \hbox{ \strut + Stability + Testability = 173}}}                                                               	\\ \hline
\end{tabular}
  \label{tab:94}
\end{table}

 \begin{table}[htpb]
    
  \renewcommand{\arraystretch}{1.3}
\caption{Version 2 Metrics$'$ Values}
\begin{tabular}{|l|l|l|l|l|}
\hline
\multicolumn{1}{|c|}{\multirow{2}{*}{Version}} & \multicolumn{1}{c|}{\multirow{2}{*}{Metric}} & \multirow{2}{*}{Value} & \multicolumn{2}{l|}{Range of acceptable Value} \\ \cline{4-5}
\multicolumn{1}{|c|}{}                     	& \multicolumn{1}{c|}{}                    	&                    	& Min                	& Max               	\\ \hline
\multirow{11}{*}{Ver2}                     	& M\_cl\_comf2                             	& 70                 	& 0                  	& 100               	\\ \cline{2-5}
                                           	& M\_cl\_data2                             	& 14                 	& 0                  	& 25                	\\ \cline{2-5}
                                           	& M\_cl\_data\_publ2                       	& 3                  	& 0                  	& 7                 	\\ \cline{2-5}
                                           	& M\_cl\_func2                             	& 2                  	& 0                  	& 9                 	\\ \cline{2-5}
                                           	& M\_cl\_func\_publ2                       	& 8                  	& 0                  	& 20                	\\ \cline{2-5}
                                           	& M\_cl\_stat2                             	& 2                  	& 0                  	& 7                 	\\ \cline{2-5}
                                           	& M\_cl\_wmc2                              	& 6                  	& 0                  	& 11                	\\ \cline{2-5}
                                           	& M\_cu\_cdused2                           	& 3                  	& 0                  	& 6                 	\\ \cline{2-5}
                                           	& M\_cu\_cdusers2                          	& 1                  	& 0                  	& 3                 	\\ \cline{2-5}
                                           	& M\_in\_bases2                            	& 1                  	& 0                  	& 4                 	\\ \cline{2-5}
                                           	& M\_in\_noc2                              	& 2                  	& 0                  	& 5                 	\\ \hline
\multicolumn{5}{|l|}{\vtop{\hbox{ \strut Analyzability = M\_cl\_wmc + M\_cl\_comf + M\_in\_bases} \hbox{ \strut  + M\_cu\_cdused = 6+70+1+3=80}   }}                                                      	\\ \hline
\multicolumn{5}{|l|}{Changeability = M\_cl\_stat + M\_cl\_func + M\_cl\_data =18}                                                                                   	\\ \hline
\multicolumn{5}{|l|}{\vtop{\hbox{ \strut Stability = M\_cl\_data\_publ + M\_cu\_cdusers + M\_in\_noc} \hbox{ \strut  + M\_cl\_func\_publ =14}}}                                                           	\\ \hline
\multicolumn{5}{|l|}{Testability = M\_cl\_wmc + M\_cl\_func + M\_cu\_cdused = 11}                                                                                   	\\ \hline
\multicolumn{5}{|l|}{\vtop{\hbox{ \strut Maintainability = Analyzability + Changeability} \hbox{ \strut  + Stability + Testability = 123}}}                                                               	\\ \hline
\end{tabular}
  \label{tab:93}
\end{table}



Table \ref{tab:94} and \ref{tab:93} : shows that all metrics values are within the range of acceptable values, and the values of metrics of Ver1 have greater values than values of metrics of Ver2. So, they must satisfy them as following:

\begin{itemize} 
\item For Ver1, the value of $(M\_cl\_comf1) >$ the value of $(M\_cl\_comf2)$ of Ver2.
\item For Ver1, the value of $(M\_cl\_data1) > $the value of $(M\_cl\_data2)$ of Ver2. 
\item For Ver1, the value of $(M\_cl\_data\_publ1) >$ the value of $(M\_cl\_data\_publ2)$ of Ver2. 
\item For Ver1, the value of $(M\_cl\_func1) > $the value of $(M\_cl\_func2)$ of Ver2.
\item For Ver1, the value of $(M\_cl\_func\_publ1) >$ the value of $(M\_cl\_func\_publ2)$ of Ver2 .
\item For Ver1, the value of $(M\_cl\_stat1) >$ the value of $(M\_cl\_stat2)$ of Ver2. 
\item For Ver1, the value of $(M\_cl\_wmc1) >$ the value of $(M\_cl\_wmc2)$ of Ver2. 
\item For Ver1, the value of $(M\_cu\_cdused1) > $ the value of $(M\_cu\_cdused2)$ of Ver2. 
\item For Ver1, the value of $(M\_cu\_cdusers1) >$ the value of $(M\_cu\_cdusers2)$ of Ver2. 
\item For Ver1, the value of $(M\_in\_bases1) >$ the value of $(M\_in\_bases2)$ of Ver2. 
\item For Ver1, the value of $(M\_in\_noc1) >$ the value of $(M\_in\_noc2)$ of Ver2. 

\end{itemize}

So, we can deduce that the value of Maintainability for Ver1 has a greater value than the value of Maintainability for Ver2 and we can present it mathematically as following:

	\begin{itemize} 
\item For Ver1, the value of Maintainability $>$ the value of Maintainability of Ver2. So $173 > 123$.

\end{itemize}

Therefore, we use theoretical validation approach to prove each of tracking and consistency at different time and the direction of change should be in the same for both.
 So that: IF (value of metrics change) $ \implies$ (value at criteria and factor level will change).

Finally, we can conclude from the results above that the value of Maintainability for Ver2 is better than Ver1 because the value of size of Maintainability for M Ver2 is less than the value of size of Maintainability for Ver1.


\subsubsection{Logiscope Results (GIPSY)}
Figure 3,4,5,6 and 7 shows the Maintainability and its$'$ criterias graphs for GIPSY.

\subsubsection{Class Metric Level for any two chosen classes}
We have chosen two classes from GIPSY and the first class is GIPSY.apps.memocode.genome.align.java,which is available in excellent section in changeability in classes’ category. Figure 8, shows the Kiviat Diagram for this class. The second class gipsy.Gee.config.java, which is available in excellent section in changeability in classes’ category. Figure 9, shows the Kiviat Diagram for this class.

GIPSY$'$s maintainability factor has been evaluated at different levels by Logiscope.
As a whole, the maintainability factor has been evaluated to be good or excellent for around 85\% of the entire source code (56\% as excellent and 26\% as good.)
In general we can state that GIPSY$'$s source code is quite easy to be maintained.
Since the maintainability factor is composed of lower level criteria, the analysis of these criteria would correlate with the statement provided above.
As a reminder, the maintainability factor is composed of four criteria : analyzability, changeability, stability and testability.

First, analyzability criteria measures how well the software can be analyzed in situations such as locating a failure or locate a software module where to implement a new specification.
According to the related Pie Graph, analyzability is good or excellent for 83\% of the source code, meaning that the software is proned to be analyzed easily by a developer. Second, changeability is defined as the degree to which a software is easy to be modified.
The related pie graph shows that the 91\% of the software code is evaluated to be excellent or good in terms of changeability. In other terms, whether It is quite easy to detect elements to change in the software or the specifications for changes are pretty close to the required changes or the changes doesn’t affect that much the rest of the software. Third, stability in code is achieved when few unexpected behaviors occurs as a result of changes. It is measured that 88\% of the code is highly stable.

Finally, the testability is measured as being how software modules can be written in such a way that it makes them easy to test. Measures done through Logiscope indicate that around 93\% of the software is highly testable.
In the light of what have been measured by these criteria, the software is easy to maintain since it is proned to be analyzed, easily changed, highly stable and testable. Kiviat Graphs for two chosen classes have been provided thanks to Logiscope tool.A quite number of indicators are shown by the Kiviat graph regarding GIPSY.apps.memocode.genome.align class. align class has 6 methods, all public $(cl\_func, cl\_func_publ)$.  It weighted methods per class metric (WMC $(cl\_wmc)$) is fairly low indicating a low complexity of the class, which contributes to its understandability. The low complexity statement is confirmed by the low number of statements ($cl\_stat$ only 12), the number of children ($in\_noc$ having 2 children) and number of base class $(in\_bases equals 1)$. Finally, 26\% of the code is commented which shows that the class is highly documented which increases its understandability.  All the aspects mentioned above regarding the align class relate a high understandable class which contributes significantly to its changeability. A class highly understandable is well proned to be changed easily. Another Kiviat has been provided regarding GIPSY.Gee.config class. Same measures has been taken as for the above mentioned class.  since the class in question does contain any methods, it is clearly a config class as it is suggested by its naming. Also, a red flag indicator is the absence of statements and only few attributes that represents different configurations. This class has no particular logic and only holds data information. The Number of children (NOC) metric equal to 6 indicates that the logic is most likely contained in the children class. These children classes overrides the configurations in the config base class for different contexts. Unfortunately, this class is not really representative of a highly changeability due to his limited amount of information.

\begin{figure} [ht!]
  \centering
    \includegraphics[width=0.5\textwidth]{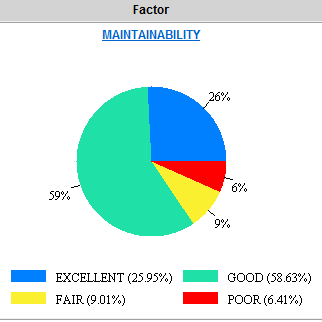}
\caption{GIPSY: Class Factor Level : Maintainability}
\end{figure}

\begin{figure} [ht!]
  \centering
    \includegraphics[width=0.5\textwidth]{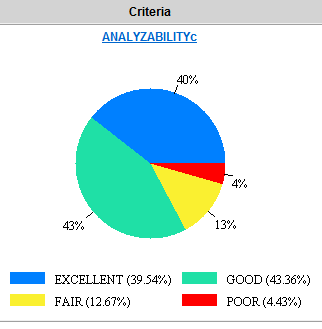}
\caption{GIPSY: Class Criteria Level: Analyzability}
\end{figure}

\begin{figure} [ht!]
  \centering
    \includegraphics[width=0.5\textwidth]{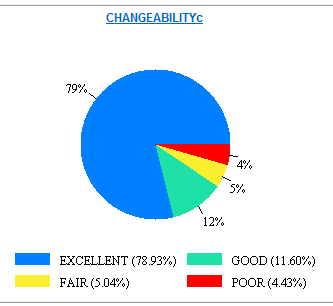}
\caption{GIPSY: Class Criteria Level: Changeability}
\end{figure}

\begin{figure} [ht!]
  \centering
    \includegraphics[width=0.5\textwidth]{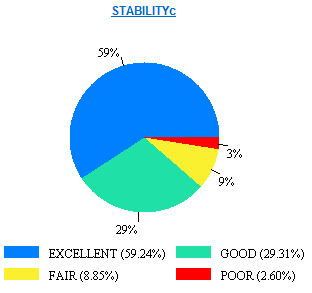}
\caption{GIPSY: Class Criteria Level: Stability}
\end{figure}

\begin{figure} [ht!]
  \centering
    \includegraphics[width=0.5\textwidth]{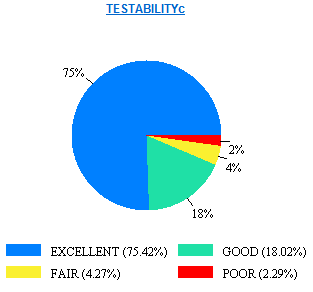}
\caption{GIPSY: Class Criteria Level: Testability}
\end{figure}

\begin{figure} [ht!]
  \centering
    \includegraphics[width=0.5\textwidth]{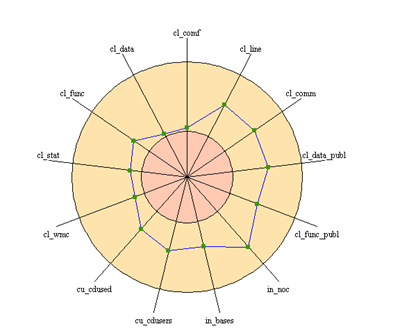}
\caption{GIPSY: Kiviat Diagram For gipsy.apps.memocode.genome.Align}
\end{figure}

\begin{figure} [ht!]
  \centering
    \includegraphics[width=0.4\textwidth]{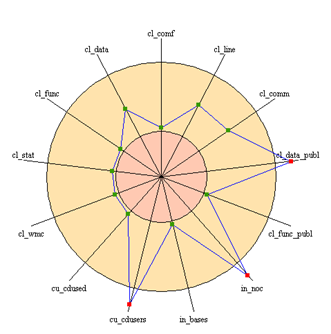}
\caption{GIPSY: Kiviat Diagram For gipsy.GEE.CONFIG}
\end{figure}

\subsubsection{Logiscope Results (MARF)}
Figure 10, 11, 12, 13 and 14 shows the Maintainability and its$'$ criterias graphs for MARF. 
 
\subsubsection{Class Metric Level for any two chosen classes}
We have chosen two classes from MARF, and the first class is Marf.gui.guiexception.java class, which is available in excellent section in CHANGEABILITY in classes’ category. Figure ~\ref{fig:42} , shows the Kiviat Diagram for this class. The second class is marf.gui.util.BorderPanel, which is available in excellent section in CHANGEABILITY in classes’ category.Figure ~\ref{fig:43}, shows the Kiviat Diagram for this class.

The overall maintainability factor of MARF is about 84\% good or excellent (having 24\% as excellent and 60\% as good). This is an indicator that in general MARF$'$s source code is easy to maintain especially that only about 6\% of the code is poor whereas the remaining 10\% fair code can be transformed to good when some effort is put on it.
As it was mentioned earlier, Maintainability factor is affected by four other lower criteria factors (Analyzability, Changeability, Stability,  Testability)
Analyzability: When it comes Analyzability criteria, we find that MARF has a minor criteria as Fair or Poor (13.5\% in total) and out of this minority only 1.4\% of the code is poor to be analyzed which indicates that MARF is easy to be analyzed in general.

{\it{\bf Changeability:}} 93\% of the MARF source code is indicated to have an excellent or good changeability criteria and only 1.85\% of the code has poor changeability. This indicates that the software is very easy to be changed later by developers.

{\it{\bf Stability:}}20\% of the code has fair or poor stability with only 3.7\% has poor stability and the rest 80

{\it{\bf Testability:}}As it is clearly indicated by the testability criteria pie chart, MARF has an excellent or good testability of 94\%. This 94\% clearly states that MARF is a highly testable software and it can be easily tested .

Again, since maintainability is a function of Analyzability, Changeability, Stability,  Testability $(Maintainability = Analyzability + Changeability +Stability + Testability)$ and as it was stated before (after analyzing the pie charts generated by Logiscope) the high analyzability, changeability, stability, and testability contributes to a highly maintainable software as it was also stated by the pie chart analysis previously.

\begin{figure} [ht!]
  \centering
    \includegraphics[width=0.5\textwidth]{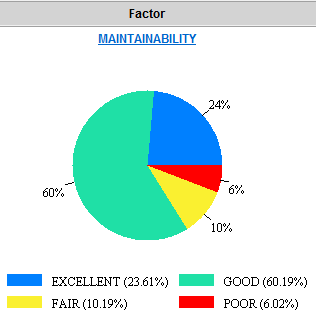}
\caption{MARF: Class Factor Level : Maintainability}
\end{figure}

\begin{figure} [ht!]
  \centering
    \includegraphics[width=0.5\textwidth]{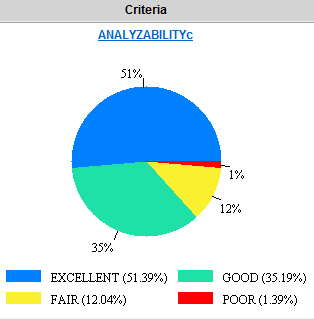}
\caption{MARF: Class Criteria Level: Analyzability}
\end{figure}

\begin{figure} [ht!]
  \centering
    \includegraphics[width=0.5\textwidth]{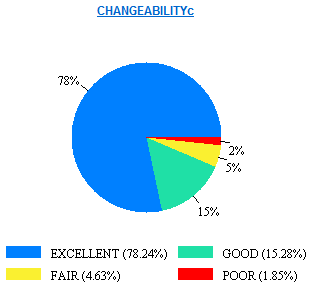}
\caption{MARF: Class Criteria Level: Changeability}
\end{figure}

\begin{figure} [ht!]
  \centering
    \includegraphics[width=0.5\textwidth]{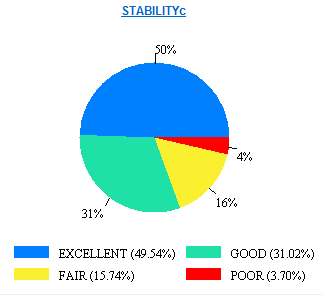}
\caption{MARF: Class Criteria Level: Stability}
\end{figure}
\begin{figure} [ht!]
  \centering
    \includegraphics[width=0.5\textwidth]{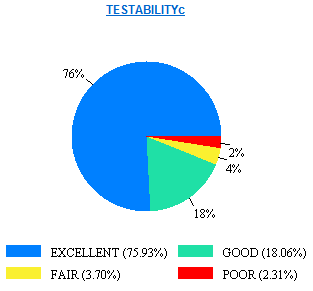}
\caption{MARF: Class Criteria Level: Testability}
\end{figure}

\begin{figure} [ht!]
  \centering
    \includegraphics[width=0.4\textwidth]{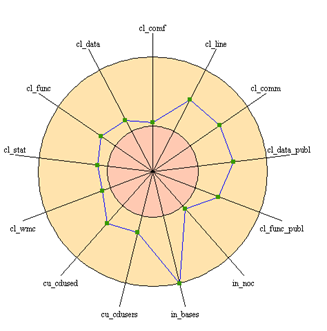}
\caption{MARF: Kiviat Diagram For Marf.gui.guiexception}
\label{fig:42}

\end{figure}

\begin{figure} [ht!]
  \centering
    \includegraphics[width=0.4\textwidth]{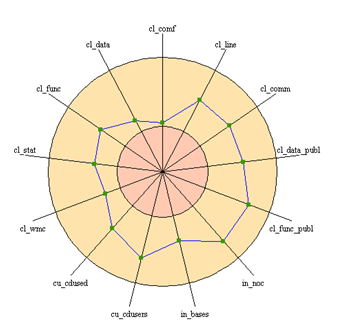}
\caption{MARF: Kiviat Diagram For marf.gui.util.BorderPanel}
\label{fig:43}

\end{figure}

{\it{\bf Kiviat Diagrams:}}

{\it{\bf marf.gui.guiexception.java class:}} Figure 15 shows the class readings show that all measures are in normal and good ranges (All of the measured attributes fall between the minimum and maximum interval) and all of them have the ‘status’ indicator as zero (0) which also implies that the there is no issue with the measured attribute (usually -1 indicated the existence of an out of range measurement).

	\begin{itemize} 
\item We start with the Class comment rate: the readings show that about 66\% of the class code is commented, which contributes to high readability and understandability.
\item Total number of attributes and number of public attributes: even though the class has only one attribute, this attribute is not public implying that the class has a high AHF (attribute hiding factor).

\item Methods: This class has 5 methods in total and all are public. Its weighted methods per class metric is 5 (and the maximum is 60). This shows that the class’s WMC is relatively low and it has low complexity.

\item Number of statements: the class has a relatively low number of statements (10 and the maximum is 100) which also contributes to low class complexity.

\item The low number of direct used/users classes, number of base classes, and number of children also contributes to the above mentioned low complexity and understandability.

\item From the above mentioned measurements we can conclude that the class marf.gui.guiexception.java has a high changeability tendency due to the low complexity contributing in high understandability and readability of the class.

\end{itemize}

{\it{\bf  marf.gui.util.BorderPanel:}} Figure 16 shows the class readings show that all measures are also in normal and good ranges and all of them have the ‘status’ indicator as zero (0).

\begin{itemize} 
\item Class comment rate: the readings show that about 50\% of the class code is commented, which contributes to high readability and understandability as mentioned previously.

\item Total number of attributes and number of public attributes: This class has only one attribute and this attribute is not public implying that the class has a high AHF.

\item
Methods: This class has 10 methods in total and all are public. Its weighted methods per class metric is 11 (and the maximum is 60). This shows that the class’s WMC is relatively low and it has low complexity. 

\item
Number of statements: the class has a relatively good number of statements (30) which also contributes to relatively low class complexity.

\item The low number of direct used/users classes, number of base classes, and number of children also contributes to the above mentioned low complexity and understandability.

\item
From the above mentioned measurements we can conclude that the class marf.gui.util.BorderPanel.java has a high changeability tendency due to the low complexity contributing in high understandability and readability of the class.

\end{itemize}

\subsubsection{Analyzing the collected measurement data}

 {\it{\bf  Comparison between the maintainability:}}
Table \ref{tab:8}, shows the obtained results from LOGISCOPE for both Open Software System OSS as follows. In the class factor level, especially in the fair category, we found that value of maintainability of MARF is higher than the value of maintainability of GIPSY. However, in the poor category we observed that value of maintainability of MARF is less than the value of maintainability of GIPSY. 

In the criteria factor level, especially in the poor category, we noticed that values of analyzability, changeability of GIPSY is higher than the values of analyzability, changeability of MARF. So, that mean the GIPSY is more complex and difficult to comprehend than MARF. While, values of testability of GIPSY considers close to the value of testability of MARF. However, a value of Stability of GIPSY is less than the value of Stability of MARF. For the criteria level, particularly in the fair category, we found that values of analyzability, changeability of GIPSY is higher than the values of analyzability, changeability of MARF. So, that mean the GIPSY is more complex and difficult to comprehend than MARF. We deduce from obtained results that the value of maintainability of GIPSY is worse than the value of maintainability of MARF based on the poor and fair category of the maintainability.

\begin{table}[h]
 \renewcommand{\arraystretch}{1.3}
\caption{LOGISCOPE Results for both Open Software System OSS}

\begin{tabular}{|l|l|l|l|l|l|}

\hline
OSS                	&             	& Excellent & Good  & Fair  & Poor  \\ \hline
\multirow{5}{*}{MARF}  & Maintainability & 23.61 	& 60.19 & 10.19 & 6.02  \\ \cline{2-6}
                   	& Analyzability   & 51.39 	& 35.19 & 12.04 & 1.39  \\ \cline{2-6}
                   	& Changeability   & 78.24 	& 15.28 & 4.63  & 1.85  \\ \cline{2-6}
                   	& Stability   	& 49.54 	& 31.02 & 15.74 & 3.70  \\ \cline{2-6}
                   	& Testability 	& 75.93 	& 18.06 & 3.70  & 2.31  \\ \hline
\multirow{5}{*}{GIPSY} & Maintainability & 25.95 	& 58.63 & 9.01  & 6.415 \\ \cline{2-6}
                   	& Analyzability   & 39.54 	& 43.36 & 12.67 & 4.43  \\ \cline{2-6}
                   	& Changeability   & 78.93 	& 11.60 & 5.04  & 4.43  \\ \cline{2-6}
                   	& Stability   	& 59.24 	& 29.31 & 8.85  & 2.60  \\ \cline{2-6}
                   	& Testability 	& 75.42 	& 18.02 & 4.27  & 2.29  \\ \hline
\end{tabular}
 \label{tab:8}
\end{table}

\subsubsection{Poor and Fair Classes}
 
 For each case study, we have identifies and lists the classes, which are characterized as fair or poor.

 {\it{\bf  1- GIPSY:}}There are many classes which are characterized as fair and poor category in factor level and criteria level. Because of the space limitation,we had to place the list of Fair and Poor classes to the appendix. Where appendix A has all GIPSY$'$s Classe.

 {\it{\bf  Maintainability: }} There are a 59 of classes which is characterized as fair category as described in figure ~\ref{fig:44} , while the 43 of classes is characterized as poor category as described in figure ~\ref{fig:45} . 

 {\it{\bf Analyzability:}} There are a 83 of classes which is characterized as fair category as described in figure  ~\ref{fig:46} , while the 29 of classes is characterized as poor category as described in figure  ~\ref{fig:47} .

 {\it{\bf Changeability: }} There are a 33 of classes which is characterized as fair category as described in figure ~\ref{fig:48} , while the 29 of classes is characterized as poor category as described in figure ~\ref{fig:49} .

 {\it{\bf Stability:}} There are a 58 of classes which is characterized as fair category as described in figure ~\ref{fig:50} , while the 17 of classes is characterized as poor category as described in figure ~\ref{fig:51}.

{\it{\bf Testability: }}There are a 28 of classes which is characterized as fair category as described in figure ~\ref{fig:52} , while the 15 of classes is characterized as poor category as described in figure ~\ref{fig:53} .

{\it{\bf  2- MARF:}} There are many classes which are characterized as fair and poor category in factor level and criteria level.Because of the space limitation,we had to place the list of Fair and Poor classes to the appendix B that has all MARF’s Clasess.

{\it{\bf  Maintainability: }} There are a 22 of classes which is characterized as fair category as described in figure ~\ref{fig:57} , while the 13 of classes is characterized as poor category as described in figure ~\ref{fig:58} .

{\it{\bf  Analyzability:}} There are a 26 of classes which is characterized as fair category as described in figure ~\ref{fig:59} , while the 3 of classes is characterized as poor category as described in figure ~\ref{fig:60} .

{\it{\bf Changeability:}} There are a 10 of classes which is characterized as fair category as described in figure ~\ref{fig:61} , while the 4 of classes is characterized as poor category as described in figure ~\ref{fig:62} .

{\it{\bf  Stability:}} There are a 34 of classes which is characterized as fair category as described in figure ~\ref{fig:63} , while the 8 of classes is characterized as poor category as described in figure ~\ref{fig:64} .

Testability: There are a 8 of classes which is characterized as fair category as described in figure ~\ref{fig:65} , while the 8 of classes is characterized as poor category as described in figure ~\ref{fig:66} .

\subsubsection{Comparing the quality of two classes}

\begin{figure} [ht!]
  \centering
    \includegraphics[width=0.5\textwidth]{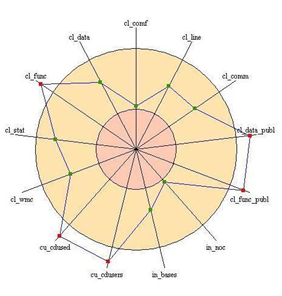}
\caption{GIPSY: Kiviat Diagram For gipsy.Configuration}
\label{fig:54}

\end{figure}

\begin{figure} [ht!]
  \centering
    \includegraphics[width=0.5\textwidth]{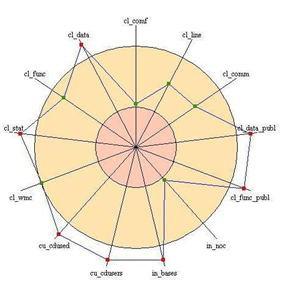}
\caption{GIPSY: Kiviat Diagram For gipsy.GIPC.GIPC}
\label{fig:55}

\end{figure}

{\it{\bf Comparison Between gipsy.Configuration Class and gipsy.GIPC.GIPC:}} 
\begin{itemize} 
\item    In GIPC class, we observed that the total number of attributes declared is 50, which is much higher than the threshold specified in Logiscope, which is 7. However, it is 4 in Configuration class, which is in an acceptable range.
    
\item    In both classes, it is clear that the number of public attributes is not acceptable. Especially, in GIPC class which has 36 public attributes. Therefore, it has tremendously exceeded the maximum range. While Configuration class has only 2 public attributes.

\item        In Configuration class, we observed that the total number of methods is 27, which is slightly over the acceptable threshold, while in GIPC they are 19, which is acceptable. 

\item          In GIPC class, we found that the number of public methods is 16, which is slightly over the maximum threshold. However, the Configuration class has 26, which means that it has tremendously exceeded the maximum range.

\item         In GIPC, we found that the number of statements is 233, which is tremendously higher than the maximum threshold. Nevertheless, Configuration class has 66, and did not surpassed the threshold.

\item          In both classes, it is clear that the number of direct used classes is over the maximum range. GIPC class has 45, and Configuration class has 12.

\item         The number of used classes is 12, which is slightly higher the threshold, while the number of user class is 10 time higher than the acceptable range in Configuration class. In contrast, the used classes is 4 times higher than the acceptable range, and the number of user class is a little bit above the acceptable range.  

\item         All other parameters are in the accepted range.
\end{itemize}

\subsubsection{{\bf Recommendations}}
{\it {\bf Recommendations at class factor/criteria level:}}
Table \ref{tab:9} , shows the recommendations for the case study at class factor/criteria level for both case study.

 \begin{table}[htpb]
    
  \renewcommand{\arraystretch}{1.3}
 \caption{Recommenations at class factor/criteria level}

 \begin{tabular}{|c|p{5cm}|} 
 
\hline
Case Study         	& Recommenations                                                                                                                                                                                                                                                    	\\ \hline
\multirow{5}{*}{GIPSY} & To improve the analyzability in the 83 fair and 29 poor classes, we recommend increasing the number of comment lines to improve the (cl\_comf), in addition to this decrease the number of direct used classes (cu\_cdusers).                                     	\\ \cline{2-2}
                   	& To improve the changeability in the 33 fair and 29 poor classes, the total number of attributes and number of methods declared should be decreased.                                                                                                               	\\ \cline{2-2}
                   	& To improve the stability, we recommend decreasing the number of public methods (cl\_func\_publ), decrease Number of direct users classes (cu\_cdusers), and decrease Total number of attributes (cl\_data). These should be done for the 58 fair and 17 poor classes. \\ \cline{2-2}
                   	& Improving the Testability can be achieved by reducing the number of methods and number of used classes for the 28 fair and 15 poor classes.                                                                                                                       	\\ \cline{2-2}
                   	& Finally, implementing the above recommendations will result in improving the overall maintainability quality of GIPSY                                                                                                                                             	\\ \hline
\multirow{5}{*}{MARF}  & To improve the analyzability in the 26 fair classes and 3 poor classes, it is recommended to reduce the number of direct users class improve the (cu\_cdused).                                                                                                    	\\ \cline{2-2}
                   	& To improve the changeability in the 10 fair and 4 poor classes, it is important to refactor the classes that have high number of attributes or methods, and may split some classes into multiple classes.                                                         	\\ \cline{2-2}
                   	& To improve the stability, it is recommended to declare all attributes as private and declare only the accessible method to be public, otherwise they should be private or protected as needed to decrease the number of public methods (cl\_func\_publ).          	\\ \cline{2-2}
                   	& Improving the Testability can be achieved by reducing the number of methods and number of used classes for the 8 fair and 8 poor classes.                                                                                                                         	\\ \cline{2-2}
                   	& \begin{tabular}[c]{@{}l@{}}Finally, applying the mentioned \\ recommendation should improve the \\ overall maintainability quality of MARF\end{tabular}                                                                                                               	\\ \hline
\end{tabular}
\label{tab:9}
\end{table}

{\it {\bf Recommendations at metric level for the two chosen classes:}}

\begin{itemize} 
\item    All attributes should be private and be accessed through their getter and setter methods. Therefore, the encapsulation of the classes will be improved.

\item
To be able to reduce the coupling of both classes, it is recommended to split each one of them into two or more classes, or even a subclass that inherits the shared attributes and methods.  As result, the number of classes that directly use gipsy.Configuration can be decreased by introducing a subclasses, while decrease the number of classes that directly use gipsy.GIPC.GIPC by splitting the class into two or more classes. 
\end{itemize}

\subsection{McCabe}

\subsubsection{Analyzing the quality trends of methods}

 \begin{table}[htpb]
    
  \renewcommand{\arraystretch}{1.3}
 \caption{Quality Trends of MARF$'$s Methods}
 
 \begin{tabular}{|c|c|c|} 

    \hline
  Quality trends of methods (Modules)  & GIPSY & MARF \\ \hline

Average Cyclomatic complexity [v(G)] &  4.07 & 1.75  \\ \hline

Essential Complexity [ev(g)] &  1.84 & 1.20  \\ \hline

Module design complexity [iv(G)] & 3.01 &
1.57
 \\ \hline
 
 \end{tabular}
  \label{tab:10}
 \end{table}

We ran McCabe IQ tool on our case study projects. The results are shown in Table \ref{tab:10} which illustrates averages of three quality measures Cyclomatic Complexity, Essential Complexity and Module Design Complexity. In terms of average Cyclomatic complexity, GIPSY system has significantly higher value compared to MARF. That means GIPSY is more complex than MARF. However, this can be due to the fact that GIPSY has more classes than MARF. Second, McCabe Essential Complexity shows that the two case study projects have relatively the same value. It is clear that GIPSY and MARF have a good quality in terms of code structuredness and overall code quality. The last measure that is shown in Table \ref{tab:10} is Module Design Complexity. The values of GIPSY and MARF are 3.01 and 1.57, respectively. Finally, these measurements show a well-designed software with good quality of code, which indicates that a feasible maintenance can be conducted on these projects.

\subsubsection{Analyzing the quality trends of classes}

\begin{table}[h]
\renewcommand{\arraystretch}{1.3}
 \caption{Quality Trends of MARF$'$s Classes}
 
\begin{tabular}{|l|l|r|}
\hline
\multicolumn{1}{|c|}{Quality trends of Classes} & GIPSY & MARF  \\ \hline
Average Coupling between Object {[}CBO{]}       & 0.07  & 0.17  \\ \hline
Weighted Methods Per Class {[}WMC{]}            & 10.54 & 11.41 \\ \hline
Response for Class {[}RFC{]}                    & 12.62 & 16.43 \\ \hline
Depth of Inheritance Tree {[}DIT/Depth{]}       & 2.02  & 2.14  \\ \hline
Number of Children {[}NOC{]}                    & 0.21  & 0.25  \\ \hline
\end{tabular}
\label{tab:11}
\end{table}

To measure the quality of the classes we have collected the result from McCabe, as shown in Table \ref{tab:11}. First, the table shows the average coupling between objects for both GIPSY and MARF. CBO of MARF is higher than GIPSY$'$s which indicates that MARF is more sensitive to change and more difficult to maintain. Moreover, Depth of Inheritance Tree and Number of Children for both systems are relatively equal. DIT/Depth of GIPSY is 2.02 while MARF is 2.14. These measures shows that MARF is more difficult to understand, test, and maintain than GIPSY since its DIT/Depth is higher than MARF. The concept of reuse quality is shown in both systems, since they have equal value of NOC. The overall quality of the systems are measured using weighted methods Per Class which has an average values of 10.54 and 11.41 for GIPSY and MARF, respectively. Finally, the coupling of these two systems is measured using Response for a Class (RFC). Form the table, RFC of GIPSY is 12.62 and MARF is 16.43 which indicates that the coupling between classes of both system is high.

\subsection{Evaluation of the Logiscope tool and its comparison with McCabe IQ tool}
Logiscope is tool that allows the evaluation of source code quality in order to reduce maintenance cost, error correction or test effort.
We work on this tool and we use only quality checker in our project. The following items are examined such as sub-program complexity, application architecture, comments, and control flow structure. There are some advantages and disadvantages from using Logiscope that are clarified in the following points:

\begin{itemize} 
\item  It helps to identify changes needed in code structure for an easier maintenance so that it helps to improve software maintainability.

\item
It is implementing ISO 9126 quality model such as maintainability that is composed of following analyzability, changeability, testability, and stability. 

\item
It can be applied directly to object code [25].
\item
It can generate all measurement data that are collecting from source code  and visualize them in graphical representations and export them to graphs. 
\end{itemize}

{\it {\bf Disadvantages:}}
\begin{itemize} 
\item It support only limited programming languages such as C, ADA, Java ,and currently you  can pay license for using C\# ,and Visual basic ,but 

\item it does support some language for web development such as PHP, ASP, etc.
\item It is unable to tell too much about how well you have covered your logic [26].
\item It supports only maintainability and reusability criteria for ISO| 9126 Quality model.
\item It needs to upgrade the Java Development Kit regularly to generate the graphs on your machine.
\end{itemize}

The McCabe IQ is tool that uses to analyze the quality of the code, increase code coverage to improve the accuracy and efficiency of testing efforts [25] , and view the structure and unstructured of the code [25]. There are some advantages from using McCabe IQ that is clarified in the following points:

\begin{itemize} 
\item 
Languages: McCabe IQ  supports  the most languages [24].
\item 
Platforms:  McCabe IQ works on ANY platform including Windows 2000/XP/Vista/7, Ubuntu Linux, Linux RedHat Enterprise, and 64 bit environments that permit running of 32 bit applications [24].

\item  Advanced Visualization: McCabe IQ can generate and form the results as graphs such as battlemap (structure chart, class diagram), attack map, attack subtree, flow graphs, scatter plots, trending reports, and more to visualize your code, architecture, and design [24].

\item  A Static Analysis Tool: McCabe IQ gives you idea about error rates, shorter testing cycles, and reduced software maintenance efforts. It gives around 100 metrics that include the McCabe Cyclomatic Complexity metric, and provides the flexibility to import and customize your own set of metrics [24].

\item A Dynamic Analysis Tool: McCabe IQ delivers the most stringent test coverage technology available - Basis Path Coverage.  It produces end-to-end unit level test conditions and end-to-end integration subtree test conditions for complete unit level and integration level test planning.  It monitors logic based code coverage and produces untested test conditions so that test effectiveness can be incrementally increased [24].
\end{itemize}

\clearpage

{\it {\bf Disadvantages:}}

\begin{itemize} 
\item 
It does not support some programming languages for web development such as PHP, ASP, etc.

\item
It supports only maintainability and reusability criteria for ISO| 9126 Quality model.
\end{itemize}

\section{Design and Implementation with JDeodorant and MARFCAT}
\subsection{Overview}

{\it {\bf JDeodorant:}}
Is an Eclipse plug-in, it follows an approach that aims at identifying refactoring opportunities which resolve bad smells existing in source code [29] . It provides a complete solution for the design problems being faced by covering all the activities of the refactoring process including the selection and application of appropriate refactoring transformations, while existing approaches focus only on the detection of design problems. Furthermore, it ensures that the extracted refactoring solutions are feasible and behavior preserving by examining a set of preconditions and rules. Additionally, it pre-evaluates the effect of the extracted refactoring solutions on design quality and provides a ranking mechanism allowing the maintainers to prioritize their effort on parts of the program that would benefit the most. The plug-in employs the ASTParser of Eclipse Java Development Tools (JDT) to analyze the source code of Java projects and the ASTRewrite to apply the refactorings and provide undo functionality. 

{\it {\bf MARFCAT:}}
A MARF-based Code Analysis Tool, was first exhibited at the Static Analysis Tool Exposition (SATE) workshop in 2010 to machine-learn from the (Common Vulnerabilities and Exposures) CVE-based vulnerable as well as synthetic CWE-based (Common Weakness Enumeration) cases to verify the fixed versions as well as non-CVE based cases from the projects written in same programming languages [30].  
Machine learning techniques along with signal and NLP processing to static source and binary code analysis are applied and employed in search to detect, classify, and report weaknesses related to vulnerabilities or bad code practices found in artificial constrained languages, such as programming languages and their compiled counterparts.

The core methodology principles include classical n-gram and smoothing techniques (add-, Witten-Bell, MLE, etc.) and machine learning combined with dynamic programming. The system is given the examples of files with known weaknesses and MARFCAT learns them by computing various language models from the CVE selected test cases. CVEs and CWEs: The CVE-selected test cases serve as a primary source of the knowledge base to gather information of how known weak code “looks like” in the language model, which they store clustered per CVE or CWE. Thus, primarily, the following is being done: (a) train the system from the CVE-based cases; (b) test on the CVE-based cases; (c) test on the non-CVE-based cases. For synthetic cases we do similarly: (a) train the system from the CWE-based synthetic cases; (b) Test on the CWE-based synthetic cases; (c) Test on the CVE and non-CVE-based cases for CWEs from synthetic cases.

{\it {\bf Tool Experience:}}
Using both tools, JDeodorant and MARFCAT, for the first time we had faced some issues and we have also noticed some good features and functionalities. Starting with JDeodorant, there is no proper documentation for the API functions, such as comments. In general JDeodorant is a useful tool which needs some enhancements, such as the abovementioned points, and to support other programming languages other than Java.
As for MARFCAT, in the beginning there was no enough information and guidelines on how to operate and run it until the lecture notes were updated and we knew how to operate it on Linux. The first thing to notice about the MARFCAT is the bad user interface (UI), which is currently command-line using terminal. After that a large log file is generated and manual checking, through the log file, for classes having highest ranking weakness/vulnerability. This process was time consuming and it may lead to an unintentional omission of classes. A solution would be extracting classes having issues in a separate log file. On the other hand MARFCAT was relatively fast in terms of code analysis and log file generation.

{\it {\bf  MARFCAT advantages and disadvantages:}}

Following is a list of advantages:
\begin{itemize} 
\item Relatively fast (e.g., Wireshark's about 2400 files train and test in about 3 minutes)[30].
\item  Language-independent (no parsing) - given enough examples can apply to any language, i.e. methodology is the same no matter C, C++, Java or any other source or binary languages are used [30].
\item  Can automatically learn a large knowledge base to test on known and unknown cases [30].
\item  Can be used to quickly pre-scan projects for further analysis by humans or other tools that do in-depth semantic analysis as a means to prioritize [30].
\item  Generally, high precision (and recall) in CVE and CWE detection, even at the file level [30].
\item  A lot of algorithms and their combinations to select the best for a particular task or class [30].
\item  Can cope with altered code or code used in other projects [30].
\end{itemize}

Following is a list of disadvantages:

\begin{itemize} 
\item  No nice GUI. Presently the application is script/command-line based.As for the advantages, the following are the key benefits of MARFACT.
\item  No Good documentation or guidelines on how to use MARFCAT.
\end{itemize}

{\it {\bf  JDeodorant advantages and disadvantages:}}
Following is a list of advantages:
\begin{itemize} 
\item JDeodorant is an excellent tool for parsing of the source code and detection of inheritance tree,
\item Excellent tool for detection of methods invocations.
\end{itemize}

 Following is a list of disadvantages:
\begin{itemize} 
\item  There is no proper documentation for the API functions, such as comments. There is only the method name that can help in finding out what does the method do.
\item  JDeodorant does not show classes that belong to the same package together, rather it shows all system classes together and manual search and arrangement of classes that belong to the same package has to be done.
\item  JDeodorant does not show all inherited methods. A list of superclasses has to be generated, then get all their respective methods, and check if each method can be inherited based on its accessibility and count it in the MOOD metrics (such as MIF).
\item  JDeodorant did not work on machines having jre6 and it required machines having jre7 and above to operate.
\item JDeodorant is language-dependent and works on Java only.
\end{itemize}

\subsection{Design and Implementation}
In addition, Team 13 demonstrates two most problematic classes for both MARF and GIPSY, which are identified in PM3 by Logiscope and McCabe. Also, we try to clarify the issues that are relevant to metrics values of each class.
For MARF, we have selected two most top problematic classes which are identified in PM 3. These classes are characterized as poor category in the maintainability factor level. These classes are called marf.MARF.java and Marf.util.Arrays.java and issues of these classes are shown in Table \ref{tab:92} and \ref{tab:91}.

\begin{table}[h]
 \caption{Metrics values of Marf.MARF.java class.}
\begin{tabular}{|c|c|c|c|c|}
\hline
Class Name                               	& \multicolumn{4}{c|}{Marf.MARF.java} \\ \hline
Metrics                                  	& Value	& Min   & Max   & Status   \\ \hline
\vtop{\hbox{ \strut  cl\_data: Total} \hbox{ \strut number of attributes }}    	& 84   	& 0 	& 7 	& -1   	\\ \hline
\vtop{\hbox{ \strut  cl\_data\_publ:} \hbox{ \strut  Number of public attributes}}  & 64   	& 0 	& 0 	& -1   	\\ \hline
\vtop{\hbox{ \strut  cL\_fun: Total}\hbox{ \strut  number of methods}}       	& 58   	& 0 	& 25	& -1   	\\ \hline
\vtop{\hbox{ \strut  cl\_func\_publ:}\hbox{ \strut  Number of public methods}}	& 55   	& 0 	& 15	& -1   	\\ \hline
\vtop{\hbox{ \strut  cl\_wmc: Weighted} \hbox{ \strut  Methods per Class}}    	& 82   	& 0 	& 60	& -1   	\\ \hline
\vtop{\hbox{ \strut  cu\_cdused: Number} \hbox{ \strut of direct used classes}}	& 26   	& 0 	& 10	& -1   	\\ \hline
\vtop{\hbox{ \strut  cu\_cdusers: Number} \hbox{ \strut of direct users classes}} & 20   	& 0 	& 5 	& -1   	\\ \hline
\end{tabular}
\label{tab:92}
\end{table}

   From Table \ref{tab:92} above, we observe that Marf.MARF.java class have some issues and these issues are summarized as follows: 1) the value of number of direct used classes is 26 (measured by cu\_cdused metric ) which exceeds acceptable region (between 0 to 10 ), while the value of number of direct users classes is 20 (measured by cu\_cdusers metric ) which passes over accepted range (between 0 to 5 ). These high values, indicate a high degree of coupling between classes and a low degree of cohesion for each class. 2) The value of Weighted Methods per Class is 82 (measured cl\_wmc metric) which goes beyond acceptable area (between 0 to 60). This high value, indicates a high degree of complexity. 3) The value of number of public attributes is 64 (measured cL\_data\_publ) which goes beyond acceptable area (between 0 to 0). This high value, indicates a low degree of information hiding (a low degree of encapsulation). 4) The value of number of public methods is 55 (measured cl\_func\_publ) which surpasses the sensible area (between 0 to 15). This high value, indicates a high degree of polymorphism and coupling.  5) The value of total number of attributes is 84 (measured by cl\_data metric) which exceeds acceptable area (between 0 to 7), whereas the value of total number of methods is 58 (measured by cL\_fun metric) which passes over accepted range (between 0 to 25). This high value, indicates a high degree of complexity of methods and a low degree of cohesion between methods.

\begin{table}[h]
 \caption{Metrics values of Marf.util.Arrays.java class.}
\begin{tabular}{|c|c|c|c|c|}
\hline
Class Name                               	& \multicolumn{4}{c|}{Marf.util.Arrays.java} \\ \hline
Metrics                                  	& Value 	& Min 	& Max 	& Status 	\\ \hline
\vtop{\hbox{ \strut cL\_fun: Total }\hbox{ \strut number of methods}}      	& 280   	& 0   	& 25  	& -1     	\\ \hline
\vtop{\hbox{ \strut cl\_func\_publ: Number } \hbox{ \strut of public methods.}}	& 279   	& 0   	& 15  	& -1     	\\ \hline
\vtop{\hbox{ \strut cl\_stat: Number} \hbox{ \strut of statements. }}         	& 711   	& 0   	& 100 	& -1     	\\ \hline
\vtop{\hbox{ \strut cl\_wmc: Weighted}\hbox{ \strut Methods per Class}}     	& 357   	& 0   	& 60  	& -1     	\\ \hline
\vtop{\hbox{ \strut cu\_cdused: Number} \hbox{ \strut of direct used classes}}	& 12    	& 0   	& 10  	& -1     	\\ \hline
\vtop{\hbox{ \strut cu\_cdusers: Number} \hbox{ \strut of direct users classes}} & 28    	& 0   	& 5   	& -1     	\\ \hline
\end{tabular}
\label{tab:91}
\end{table}

From Table \ref{tab:91} above, we have noticed that Marf.util.Arrays.java class have some issues and we have outlined these issues as follows: 1) the value of number of direct used classes is 12 (measured by cu\_cdused metric ) which exceeds acceptable region (between 0 to 10 ), while the value of number of direct users classes is 28 (measured by cu\_cdusers metric ) overpassing accepted range (between 0 to 5 ). These high values, indicate a high degree of coupling between classes and a low degree of cohesion for each class. 2) The value of Weighted Methods per Class is 357 (measured cl\_wmc metric) which goes beyond acceptable range (between 0 to 60). This high value, indicates a high degree of complexity. 3) The value of Number of statements is 711 (measured cl\_stat) which goes beyond acceptable range of values (between 0 to 100). This high value, indicates a high degree of size of class. 4) The value of number of public methods is 279 (measured cl\_func\_publ) which surpasses the sensible area (between 0 to 15). This high value, indicates a high degree of, polymorphism, and coupling.  5) The value of total number of methods is 280 (measured by cL\_fun metric) which passes over accepted range (between 0 to 25). This high values, indicates a high degree of complexity of methods and a low degree of cohesion between methods.
As for GIPSY, the top two problematic classes, which were identified in PM 3, are gipsy.Configuration.java and gipsy.GIPC.GIPC.java . These classes are characterized as poor category in the maintainability factor level and issues in these classes are shown in Table \ref{tab:90} and \ref{tab:89}.

\begin{table}[h]
 \caption{Metrics values of gipsy.Configuration.java class}
\begin{tabular}{|c|c|c|c|c|}
\hline
Class Name                               	& \multicolumn{4}{c|}{gipsy.Configuration.java} \\ \hline
Metrics                                  	& Value 	& Min 	& Max 	& Status 	\\ \hline
\vtop{\hbox{ \strut cL\_data\_publ: Number}\hbox{ \strut  of public attributes}} & 2     	& 0   	& 0   	& -1     	\\ \hline
\vtop{\hbox{ \strut cL\_fun: Total}\hbox{ \strut  number of methods}}       	& 27    	& 0   	& 25  	& -1     	\\ \hline
\vtop{\hbox{ \strut cl\_func\_publ: Number}\hbox{ \strut  of public methods}}	& 26    	& 0   	& 15  	& -1     	\\ \hline
\vtop{\hbox{ \strut cu\_cdused: Number}\hbox{ \strut of direct used classes}}	& 12    	& 0   	& 10  	& -1     	\\ \hline
\vtop{\hbox{ \strut cu\_cdusers: Number}\hbox{ \strut  of direct users classes} }& 49    	& 0   	& 5   	& -1     	\\ \hline
\end{tabular}
\label{tab:90}
\end{table}

 From Table \ref{tab:90} above, we have found that gipsy.Configuration.java class have some issues and it is summarized as follows: 1) the value of number of direct used classes is 12 (measured by cu\_cdused metric ) which exceeds acceptable region (between 0 to 10 ), while the value of number of direct users classes is 49 (measured by cu\_cdusers metric ) which passes over accepted range (between 0 to 5 ). These high values, indicate a high degree of coupling between classes and a low degree of cohesion for each class. 2) The value of number of public attributes is 2 (measured cL\_data\_publ) which goes beyond acceptable area (between 0 to 0). This high value, indicates a low degree of information hiding (a low degree of encapsulation). 3) The value of number of public methods is 26 (measured cl\_func\_publ) which surpasses the sensible area (between 0 to 15). This high value, indicates a high degree of, polymorphism, and coupling with other classes.  4) The value of total number of methods is 27 (measured by cL\_fun metric) which passes over acceptable range (between 0 to 25). This high value, indicates a high degree of complexity of methods and a low degree of cohesion between methods.

\begin{table}[h]
 \caption{Metrics values of gipsy.GIPC.GIPC.java class.}
\begin{tabular}{|c|c|c|c|c|}
\hline
Class Name                               	& \multicolumn{4}{c|}{gipsy.GIPC.GIPC.java} \\ \hline
Metrics                                  	& Value 	& Min 	& Max 	& Status	\\ \hline
\vtop{\hbox{ \strut cl\_data: Total}\hbox{ \strut number of attributes}}    	& 50    	& 0   	& 7   	& -1    	\\ \hline
\vtop{\hbox{ \strut cL\_data\_publ: Number}\hbox{ \strut of public attributes}} & 36    	& 0   	& 0   	& -1    	\\ \hline
\vtop{\hbox{ \strut cl\_func\_publ: Number}\hbox{ \strut of public methods}}	& 16    	& 0   	& 15  	& -1    	\\ \hline
\vtop{\hbox{ \strut cl\_stat: Number}\hbox{ \strut of statements}}          	& 233   	& 0   	& 100 	& -1    	\\ \hline
\vtop{\hbox{ \strut cu\_cdused: Number} \hbox{ \strut  of direct used classes}}	& 45    	& 0   	& 10  	& -1    	\\ \hline
\vtop{\hbox{ \strut cu\_cdusers: Number}\hbox{ \strut of direct users classes}} & 12    	& 0   	& 5   	& -1    	\\ \hline
\vtop{\hbox{ \strut in\_bases: Number}\hbox{ \strut of base classes}}       	& 5     	& 0   	& 3   	& -1    	\\ \hline
\end{tabular}
\label{tab:89}
\end{table}

From Table \ref{tab:89} above, we have seen that gipsy.GIPC.GIPC.java class have some issues and it is outlined as follows: 1) The value of number of direct used classes is 45 (measured by cu\_cdused metric ) which exceeds acceptable region (between 0 to 10 ), while the value of number of direct users classes is 12 (measured by cu\_cdusers metric ) which passes over accepted range (between 0 to 5 ). These high values, indicate a high degree of coupling between classes and a low degree of cohesion for each class. 2) The value of number of public attributes is 36 (measured cL\_data\_publ) which goes beyond acceptable area (between 0 to 0). This high value, indicates a low degree of information hiding (a low degree of encapsulation). 3) The value of number of public methods is 16 (measured cl\_func\_publ) which surpasses the sensible area (between 0 to 15). This high value, indicates a high degree of, polymorphism, and coupling. 4) The value of number of base classes is 5 (measured by in\_bases metric) which passes over accepted range (between 0 to 3). This high value, indicates a high degree of Inheritance with other classes. 5) The value of total number of attributes is 50 (measured by cl\_data metric) which exceeds acceptable area (between 0 to 7). 6) The value of number of statements is 233 (measured cl\_stat) which goes beyond acceptable area (between 0 to 100). This high value, indicates a high degree of size of class and a low degree of cohesion of a class.


\subsubsection{The problematic classes occupy vs. less problematic in their respective packages in MARF and GIPSY}

In this section, we show how the size of dependency of each one of problematic classes to others classes in the entire system for each metric of C\&K metrics in both case studies.

\begin{table}[h]
\caption{the size of dependency of problematic classes at system in MARF }
\begin{tabular}{|c|c|c|}
\hline
\multicolumn{3}{|c|}{The size of the dependency of each class in the entire system in MARF} \\ \hline
Metric name            & the \% of MARF class            & the \% of Arrays class           \\ \hline
LCOM                   & 0.0102 \%                       & 0 \%                             \\ \hline
NOC                    & 0 \%                            & 0 \%                             \\ \hline
RFC                    & 0.0215 \%                       & 0.0941 \%                        \\ \hline
DIT                    & 0.0025 \%                       & 0.0025 \%                        \\ \hline
CBO                    & 0.0322 \%                       & 0 \%                             \\ \hline
WMC                    & 0.0280 \%                       & 0.1355 \%                        \\ \hline
\end{tabular}
\label{tab:87}
\end{table}

Table \ref{tab:87} above, the value of the size of the dependency of MARF.java class to others classes in the entire system of MARF is as follows: the percentage of LCOM is 0.0102\%, the percentage of NOC is 0\%, the percentage of RFC is 0.0215\%, the percentage of DIT is 0.0025\%, the percentage of CBO is 0.0322\%, the percentage of WMC is 0.0280

\begin{table}[h]
\caption{The size of dependency of problematic classes at system in GIPSY.}
\begin{tabular}{|c|c|c|}
\hline
\multicolumn{3}{|c|}{The size of the dependency of each class in the entire system in GIPSY} \\ \hline
Metric name          & the \% of GIPSY class         & the \% of Configuration class         \\ \hline
LCOM                 & 0.00327\%                     & 0.00232\%                             \\ \hline
NOC                  & 0 \%                          & 0 \%                                  \\ \hline
RFC                  & 0.00258\%                     & 0.00367\%                             \\ \hline
DIT                  & 0.00169\%                     & 0.00169\%                             \\ \hline
CBO                  & 0.01204\%                     & 0.02631\%                             \\ \hline
WMC                  & 0.00309\%                     & 0.00440\%                             \\ \hline
\end{tabular}
\label{tab:86}
\end{table}

Table \ref{tab:86}, the size of the dependency of GIPC.java class to others classes in the entire system of GIPSY is as follows: the percentage of LCOM is 0.00327\%, the percentage of NOC is 0\%, the percentage of RFC is 0.00258\%, the percentage of DIT is 0.00169\%, the percentage of CBO is 0.01204\%, the percentage of WMC is 0.00309\%. While the size of the dependency of Configuration.java class is to others classes in the entire system,  as follows : the percentage of LCOM is 0.00232\%, the percentage of NOC  is 0\%, the percentage of RFC  is 0.00367\%, the percentage of DIT is 0.00169\%, the percentage of CBO is 0.02631\%, the percentage of WMC is 0.00440\%.  Therefore, GIPC.java and Configuration.java classes have low degree of coupling, complexity, and inheritance, which have limited responsibilities to others classes in system.
In addition, we illustrate how the size of dependency of each one of problematic classes to others classes in the same package for each metric of C\&K metrics in both case studies.

\begin{table}[h]
\caption{The size of dependency of problematic classes at same package in MARF.}
\begin{tabular}{|c|c|c|}
\hline
\multicolumn{3}{|c|}{The size of the dependency of each class in the same package in MARF} \\ \hline
Metric name            & the \% of MARF class           & the \% of Arrays class           \\ \hline
LCOM                   & 0.36 \%                        & 0 \%                             \\ \hline
NOC                    & 0 \%                           & 0 \%                             \\ \hline
RFC                    & 0.55\%                         & 0.45\%                           \\ \hline
DIT                    & 0.25\%                         & 0.027 \%                         \\ \hline
CBO                    & 100 \%                         & 0 \%                             \\ \hline
WMC                    & 0.52 \%                        & 0.59 \%                          \\ \hline
\end{tabular}
\label{tab:85}
\end{table}

Table \ref{tab:85}, the value of the size of  the dependency of MARF.java class  to others classes in the same package of MARF ,  as follows : the percentage of LCOM is 0.36\%,  , the percentage of NOC  is 0\%, the percentage of RFC  is 0.55\% , the percentage of DIT is 0,25\% , the percentage of CBO is 100\%, the percentage of WMC is 0,52. While the size of the dependency of Arrays.java class to others classes in the entire system is as follows: the percentage of LCOM is 0\%, the percentage of NOC is 0\%, the percentage of RFC is 0.45\%, the percentage of DIT is 0.027\%, the percentage of CBO is 0\%, and the percentage of WMC is 0.59\%.  Therefore, MARF.java and Arrays.java classes have high degree of coupling, complexity, and low degree of inheritance, which means more likely to be difficult to understand.

\begin{table}[h]
\caption{The size of dependency of problematic classes at same package in GIPSY.}
\begin{tabular}{|c|c|c|}
\hline
\multicolumn{3}{|c|}{The size of the dependency of each class in the same package in GIPSY} \\ \hline
Metric name         & the \% of GIPSY class         & the \% of Configuration class         \\ \hline
LCOM                & 0.30 \%                       & 0.88 \%                               \\ \hline
NOC                 & 0 \%                          & 0 \%                                  \\ \hline
RFC                 & 0.258\%                       & 0.81 \%                               \\ \hline
DIT                 & 0.22\%                        & 0.50 \%                               \\ \hline
CBO                 & 100 \%                        & 100 \%                                \\ \hline
WMC                 & 0.30 \%                       & 0.81 \%                               \\ \hline
\end{tabular}
\label{tab:84}
\end{table}

Table \ref{tab:84}, the size of the dependency of GIPC.java class to others classes in the same package of GIPSY is as follows: the percentage of LCOM is 0.30 


\subsection{Implementation of Metrics with JDeodorant}

In this section, we illustrate a set of metrics have been implemented by team 13, graduate students at Concordia University by using Java programming language and API functions of JDeodorant tool, and tested by using dummy test. These metrics are nine metrics, which include three CK metrics such as Number Of Children (NOC), Depth of Inheritance Tree (DIT), Response for Class (RFC) and six MOOD metrics such as Method Inheritance Factor (MIF), Attribute Hiding Factor (AHF), Coupling Factor (CF), Method Inheritance Factor (MIF), Attribute Inheritance Factor (AIF), and  Polymorphism Factor (PF). 

\begin{figure} [ht!]
  \centering
    \includegraphics[width=0.4\textwidth]{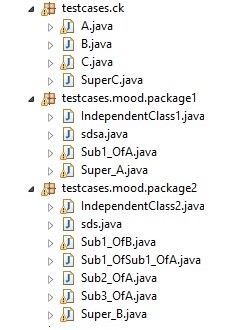}
\caption{Dummy Test Cases}
 \label{fig:70}
\end{figure}

{\it {\bf Metrics Testing:}} 
In order to test the metrics and make sure that they produce accurate results, two dummy test cases have been designed and implemented. The first test case, which is called testcases.ck package, is responsible to test RFC, DIT and NOC metrics. This test is taking from the lectures slides, where they are used to count all CK metrics. The second test case, which has two packages and they are as follow: testcases.mood.package1 package and test casses.mood.package2 package. This test case tests the MOOD metrics accuracy. Nevertheless, this test case is implemented from our understanding of how each metric works. Figure  ~\ref{fig:70} shows the test cases packages.

\subsection{Summary}

\subsection*{Results}
In this section, we demonstrate our results that are collected by using JDeodorant tool  to obtain all values of CK metrics , MOOD metrics from each of MARF and GIPSY and implemented metrics. Therefore , these results are shown as follows:

\subsubsection{Results of the implemented metrics}
Table \ref{tab:88}, illustrate all metrics values such as MHF, AHF, MIF, AIF , PF and CF that are collected by using JD to measure for each implemented metric at the project level and Table \ref{tab:88} show these metrics value.

\begin{table}[h]
 \caption{Implemented metric at the project level.}
\begin{tabular}{|c|c|}
\hline
Metrics at the Project Level       & Value    \\ \hline
MHF (Method Hiding Factor)         & 80.597\% \\ \hline
AHF (Attribute Hiding Factor)      & 94.643\% \\ \hline
MIF (Method Inheritance Factor)    & 0.0\%    \\ \hline
AIF (Attribute Inheritance Factor) & 0.0\%    \\ \hline
PF (Polymorphism Factor)           & 0.0\%    \\ \hline
CF (Coupling Factor)               & 7.576\%  \\ \hline
\end{tabular}
\label{tab:88}
\end{table}

Table \ref{tab:87}, illustrate all metrics values such as NOC, DIT, RFC, and LCOM that are collected by using JD to measure for each implemented metric and Table \ref{tab:87} show these metrics value.

\begin{table}[h]
 \caption{C\&K metrics values for each implemented metric file.}
\begin{tabular}{|c|c|c|c|c|}
\hline
\multirow{2}{*}{Classes Name} & \multicolumn{4}{c|}{C\&K Metrics} \\ \cline{2-5}
                         	& NOC	& DIT	& RFC	& LCOM   \\ \hline
metrics.MIF.java          	& 0  	& 0  	& 23 	& 16 	\\ \hline
metrics.CF.java           	& 0  	& 0  	& 13 	& 2  	\\ \hline
metrics.AIF.java          	& 0  	& 0  	& 20 	& 16 	\\ \hline
metrics.PF.java           	& 0  	& 0  	& 22 	& 16 	\\ \hline
metrics.MHF.java          	& 0  	& 0  	& 20 	& 7  	\\ \hline
metrics.LCOM.java         	& 0  	& 0  	& 10 	& 4  	\\ \hline
metrics.NOC.java          	& 0  	& 0  	& 13 	& 7  	\\ \hline
metrics.AHF.java          	& 0  	& 0  	& 14 	& 7  	\\ \hline
metrics.RFC.java          	& 0  	& 0  	& 18 	& 5  	\\ \hline
metrics.DIT.java          	& 0  	& 0  	& 18 	& 5  	\\ \hline
\end{tabular}
\label{tab:87}
\end{table}

\subsubsection{Interpretation of Results}
Nine metrics have been implemented in Java within JDeodorant application, each of the metric is represented by a single class. A quality assessment is conducted against the implemented metrics by evaluating both Mood and CK metrics.

First, the Mood metrics allow an evaluation of the quality of the code at the project level, thus giving an overall idea of the metrics package. The table above summarizing the principal results of the Mood metrics shows clearly that the attribute and method hidden factors (AHF and MHF) have a high value , in order words a high proportion of protected methods over unprotected methods, which suggests a highly specialized methods not intended for reuse. This result correlates with the nature of the project since the metrics are particularly designed for a specific purpose that is the computation of a metric. The hierarchy of the source code confirms this statement since each metric class is isolated from the rest. Since each metric class is isolated for the other, we can deduce that there is no inheritance applied, which implies null Attribute and Method Inheritance factors as well as polymorphism factor. This aspect represent a weakness in the actual design since it doesn’t benefit from the object oriented principles. Conceptually, these metrics share a common point which is being either CK metric or Mood metric , and both of these two categories are metric. If we want to bring some improvement, we would refactor the actual code by adding three superclasses CK , MOOD and Metric; MOOD class would be the superclass of the following classes : MHF, AHF, MIF, AIF , PF and CF.
Whereas, CK class would be the superclass of the following classes : LCOM, NOC, RFC and DIT.

Finally,  Metric class would be the subtype of the CK and MOOD. This refactoring would add a little bit of complexity (which would be reflected in the metrics) but would allow more reuse of the code if another metric is to be implemented. Currently, if a new metric has to be implemented, no code can be reused since it is encapsulated in isolated class. This isolation characteristic is reflected by the low value of the coupling factor which shows a weak interaction between objects. There are many advantages behind an uncoupled design : design easy to maintain by avoiding a ripple effect of change, better reusability with independent components, ...etc

Low coupling is advised but equal to a null value since the oriented object design advocates a collaborating system of objects.
Also, CK metrics have been measured for the same metrics package, allowing to have this time a class level point view of each class. As said previously, the design is characterized by the absence of inheritance and it is then obvious that the depth in inheritance and number of children factors are null for such isolated classes.

However, we note a moderate value of response for class metric (RFC) which denotes a low coupling as mentioned previously. LCOM measure has an important value for almost all the classes of the metrics package, which seems quite contradictory since each class is cohesive with the purpose of its creation. We can state that the LCOM metric gives a method point of view and is not very representative of the overall cohesiveness.

\subsubsection{C\&K metrics results of both case studies}

\begin{table}[h]
 \caption{Results of the implemented MOOD Metrics for Marf.util.Arrays.java}
\begin{tabular}{|c|c|c|c|c|}
\hline
\multirow{2}{*}{Class Name} & \multicolumn{4}{c|}{C\&K Metrics} \\ \cline{2-5}
                            & NOC    & DIT    & RFC    & LCOM   \\ \hline
Marf.MARF.java              & 0      & 0      & 71     & 1260   \\ \hline
Marf.util.Arrays.java       & 0      & 0      & 279    & 3878   \\ \hline
gisy.Configration.java      & 0      & 0      & 25     & 0      \\ \hline
gipsy.GIPSY.GIPC.java       & 0      & 1      & 29     & 77     \\ \hline
\end{tabular}
\label{tab:86}
\end{table}

According to the results shown by Table \ref{tab:86}, we note that Marf.util.Arrays.java class has a high value for response for class metric (RFC) , which is an indicator of high complexity of the class. Another aspect that could explain this  high complexity is the significant value of lack of cohesion of method (LCOM) metric that shows that the class is not cohesive and gathers lot of concerns. It is then highly advised as a refactoring to decompose this class and separate the concerns for better readability and understandability.
The same observation can be made regarding Marf.MARF.java class but at a lower degree.
As indicated by its name, the MARF class represents potentially a component with high responsibility in the MARF application, so any refactoring that would allow to delegate some responsibilities to other components helps increase the cohesion of this class and reduce its coupling thus its complexity.
At the other hand, gipsy.Configuration.java class has a null LCOM metric, which suggests a high cohesive method. It has also a moderate value of RFC metric , which confirms the claim that the configuration class is understandable and easy to maintain. Null values of number of children and Depth in inheritance metrics is another indicator of the high understandability.
Finally, gipsy.GIPC.GIPC.java class shows a moderate values of RFC and LCOM metrics relatively to previous cases. This indicates a reasonably understandable component but a significant lack of cohesion. The GIPC class has a non null but low value of the depth in inheritance (DIT) metric that would potentially decrease understandability since it takes the investigation of the superclass to fully understand its code. These factors contributes normally  in reducing the maintainability of the code. 

 \subsubsection{MOOD metrics results of both case studies}

\begin{table}[h]
\caption{Results of the implemented MOOD Metrics for GIPSY and MARF}
\begin{tabular}{|c|l|l|}
\hline
MOOD Metrics                       & MARF     & GIPSY     \\ \hline
Method Hiding Factor (MHF)         & 8.903\%  & 50.243\%  \\ \hline
Attribute Hiding Factor (AHF)      & 95.277\% & 83.912\%  \\ \hline
Method Inheritance Factor (MIF)    & 40.457\% & 26.947\%  \\ \hline
Attribute Inheritance Factor (AIF) & 47.746\% & 23.484\%  \\ \hline
Polymorphism Factor (PF)           & 86.83\% & 100\% \\ \hline
Coupling Factor (CF)               & 0.2.28\%   & 0.755\%   \\ \hline
\end{tabular}
\label{tab:85}
\end{table}

Table \ref{tab:85} shows the result of the implemented MOOD Metrics of the two case studies, GIPSY and MARF. In MOOD Metrics there are six metrics that values 0 -100\%. By understanding the above numbers, Team 13 is able to give an overall quality of the two systems. The result is divided based on the factors as follow:  

Method Hiding Factor (MHF): MARF has extremely low values of MHF with 8.903\%. This indicates a lack of information hiding which means big number of methods are unprotected. One possible reason is a lack of abstraction at the design stage. For GIPSY, the MHF is slightly above the middle which is 50.243\%. Also, GIPSY’s MHF considered to have reasonable number of specialized method that cannot be reuse.

The Attribute Hiding Factor (AHF): For this metric MARF and GIPSY have a significantly a big value by 95,277\% and 83.912\%, respectively. This indicates that most of the attributes were declared as private in both case studies. As result, it shows that systems encapsulation have ideal value.

Method Inheritance Factor (MIF): The value of MIF for MARF is 40.457\% which is relatively low percentage. This indicates a little of utilization of inheritance or heavy use of overriding in MARF. GIPSY’s MIF value is 26.947\%. This is also rather low, suggesting only a moderate use of inheritance.

The Attribute Inheritance Factor (AIF): both systems MARF and GIPSY have a low value which is 47.746\% for MARF and 23.484\% for GIPSY. These percentages are rather low, suggesting only a moderate use of inheritance. Also, they indicate overutilization of method overriding.

The Coupling Factor (CF): the coupling factor for both case studies are low. First MARF has 2.28\% percent for the whole system, and GIPSY has 0.755\%. These result shows that MARF and GIPSY have low potentially harmful side-effects such as unnecessary dependencies and limited reuse. Furthermore, the systems component can be reused and extended easily.

The Polymorphism Factor (PF): MARF’s value of PF is 86.83\% while GIPSY is 100\%. This metrics shows the polymorphism factor of system level. The two systems have high value of PF that indicate highly use of inheritance. However, the result shows that huge number of methods are overridden. in general, it is  considered suspect and might warrant further investigation.

{\it  Evaluate weak/vulnerable code of MARF and GIPSY by using MARFCAT:}
In addition to JDeodorant, we have used MARFCAT to detect  weak/vulnerable code for all files in both case studies MARF and GIPSY. we have followed all instructions that were provided in [33] to run MARFCAT tool on linux platform and then  we scan all Java files in both case studies with the provided default threshold. After that, we get two log files. First file is for gipsy and it is called  marfcat-super-fast-test-quick-gipsy-cve.  We observe that all values for warning should be reported for all java files are equal to false. while the second file is for MARF and it is called  marfcat--super-fast-test-quick-marf-cve. We note that all values for warning should be reported for all java files are equal to false.Therefore ,there are on  weak/vulnerable Java files are found.

\subsubsection{Analysis and Interpretation}
PM1 allowed to understand two given well documented projects and their respectives architectures. Having this analysis in hand, we could formulate some hypotheses or at least expectation on the quality of the code. Also, according to the nature of the projects,generally related to the object-oriented paradigm, we selected the most appropriate design factors that would express faithfully their overall quality. The essential quality factor for which our analysis is based consists of the maintainability. This last could only be measured by concrete metrics. Then, in PM3, we used some tools to extract measures that enable us to quantify the maintainability of the two projects by decomposing it into criteria. We found out thanks to LOGISCOPE, that the overall maintainability of the two projects are important. This measure seems relevant since it is at the project level and not specific to a particular component. However, Kiviat diagram focuses on particular classes, which limits the general assessment of the software quality. Finally, PM4 permits to extract values from different tools, which gives a different points of view and consolidate the general appreciation of the code. We used MARFCAT tool that allow to identify weaknesses and vulnerabilities in the entire project but doesn’t assess in a direct way the quality of the code. It can only gives a proportion of fragile component over the whole project and confirms an already formulated assessment of the code quality. Having implemented components that compute metrics using JDeodorant library, we could extract measures of MOOD and CK metrics on the two projects. The results of low coupling seem to enforce the idea of the high maintainability already stated in the previous milestone. The low coupling statement conforms to the high understandability, analyzability and thus maintainability obtained in PM3. Also, the high use of inheritance as shown by the MOOD metric results coordinates with the important value of changeability obtained in LOGISCOPE. However, some suspect values related to a polymorphism factor of 100\% and an almost null coupling factor put some doubt on the complete validity of the implementation of our metrics.
We have faced some restrictions during conducting the case study and these limitations are summarized its main points as follows:
\begin{itemize} 
 
\item Most of team member are working as full time or taken two courses in summer 1 , 2014.
\item Small team size which contains six graduate students.
\item Requirements specifications of project are not stable and many of changes are made periodically for each milestone.
\item Limited time of the project.
\end{itemize}

\section{Conclusion}

One of the most important aspects of a software quality product that shows how good is a software is its quality (such functionality, usability, reliability, efficiency, reusability, extendibility, and maintainability). To measure these external software quality attributes. Therefore, there is a need to investigate more in a set of internal software quality attributes such as coupling, complexity,  which may have strong influence on those external software quality attributes mentioned above. In addition, we need to follow an optimal mechanism that contains the best available metrics and reliable tools to achieve main goal in assess the overall quality of any software. Many metrics have been identified so far in this study such as CK metrics ( NOC, CBO, WMC, DIT, and RFC) which capture different aspects of an OO design; these metrics mainly focus on the class and the class hierarchy. It includes complexity, coupling and cohesion as well. MOOD metrics ( MHF, AHF, MIF, AIF, CF, PF) in order to estimate object oriented design and measure a system’s structural complexity, can be used to assess design quality and to guide improvement efforts. The application of object oriented design principles for example modularity and abstraction lead to better maintainability and reusability. A wide variety of object oriented metrics have been proposed to assess the testability of an object oriented system. MOOD metrics focus on measure  the quality of software product at system level which gives high level abstract about the entire quality of software product includes encapsulation, inheritance, polymorphism, and coupling. While CK metrics focus mainly on measure  the quality of software product at class level which provides more details about specific class in software system includes cohesion, inheritance, and coupling. Therefore, CK metrics are useful as quality indicators for predicting fault-prone classes. In our study, we find that most of results which are useful to assess the quality of object oriented design for both MARF and GIPSY. While results of selected problematic classes have issues with encapsulation and polymorphism , which are not cover in CK metrics. Finally , the obtained results from this study showed that the CK metrics helps development team to make better design decision regarding in object oriented design and MOOD metrics could be used to provide an overall assessment of any software system.

\bibliographystyle{IEEEtran}
\bibliography{bibliography}
%


\section*{Paper distribution}
 \begin{figure} [ht!]
  \centering
    \includegraphics[width=0.4\textwidth]{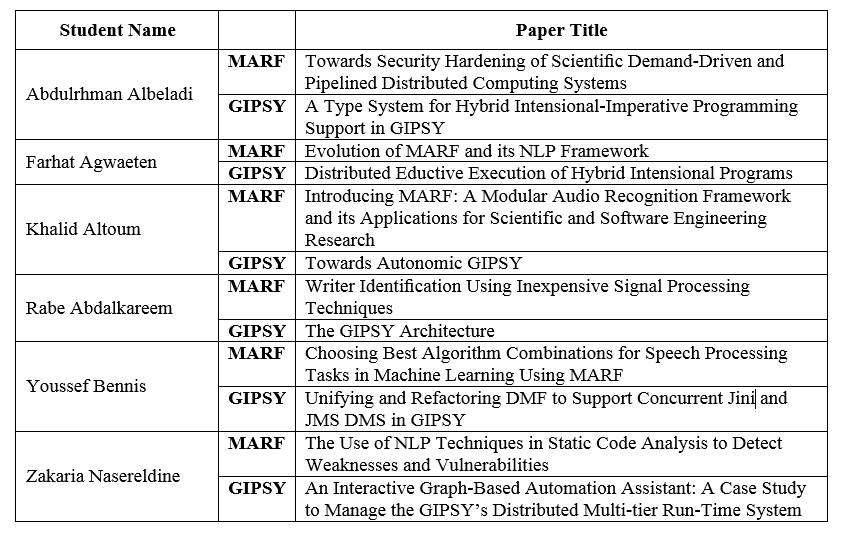}
\caption{Paper distribution: First Deliverable}

\end{figure}

 \begin{figure} [ht!]
  \centering
    \includegraphics[width=0.4\textwidth]{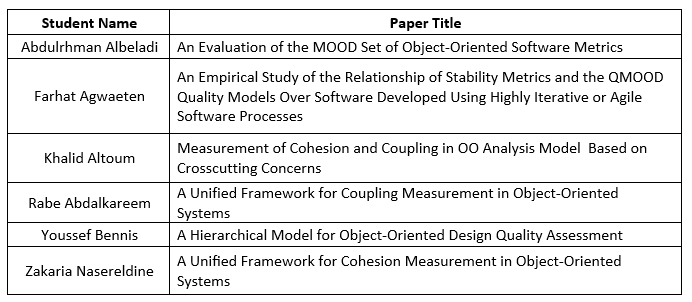}
\caption{Paper distribution: Second Deliverable}

\end{figure}

\clearpage

 
\section{ Appendix: A (GIPSY)}

\begin{figure} [ht!]
  \centering
    \includegraphics[width=0.5\textwidth]{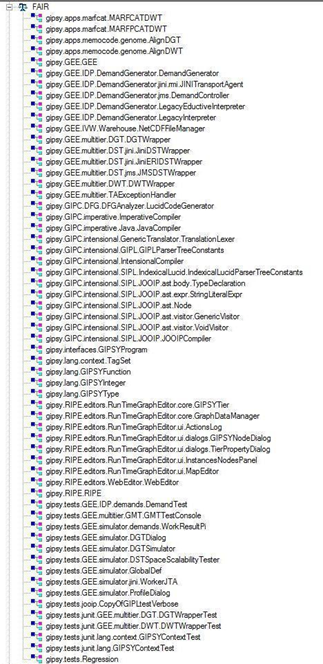}
\caption{GIPSY:Maintainability: Fair Classes}
\label{fig:44}

\end{figure}

\begin{figure} [ht!]
  \centering
    \includegraphics[width=0.5\textwidth]{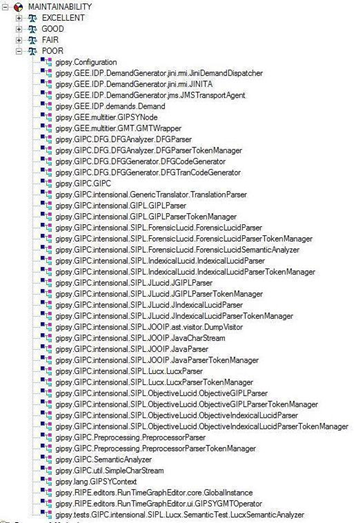}
\caption{GIPSY: Maintainability: Poor Classes}
\label{fig:45}

\end{figure}

\begin{figure} [ht!]
  \centering
    \includegraphics[width=0.5\textwidth]{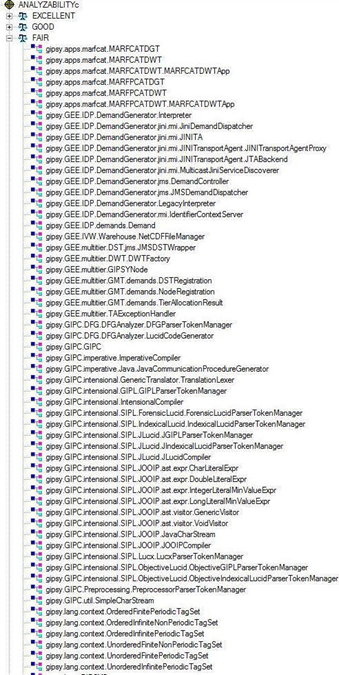}
\caption{GIPSY: Analyzability: Fair Classes}
\label{fig:46}

\end{figure}

\begin{figure} [ht!]
  \centering
    \includegraphics[width=0.5\textwidth]{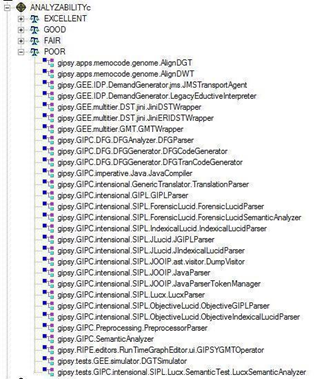}
\caption{GIPSY: Analyzability: Poor Classes}
\label{fig:47}

\end{figure}

\begin{figure} [ht!]
  \centering
    \includegraphics[width=0.5\textwidth]{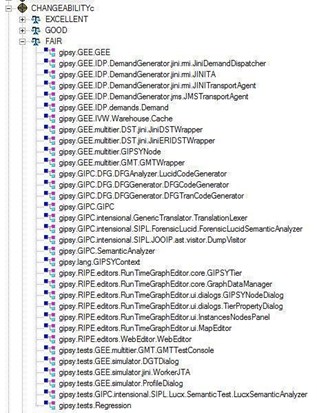}
\caption{GIPSY: Changeability: Fair Classes}
\label{fig:48}

\end{figure}

\begin{figure} [ht!]
  \centering
    \includegraphics[width=0.5\textwidth]{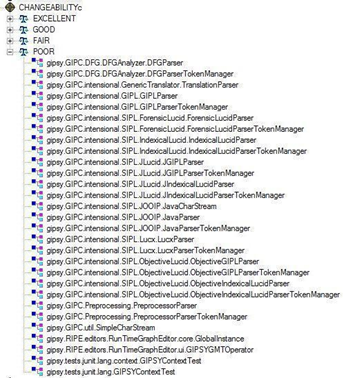}
\caption{GIPSY: Changeability: Poor Classes}
\label{fig:49}

\end{figure}

\begin{figure} [ht!]
  \centering
    \includegraphics[width=0.5\textwidth]{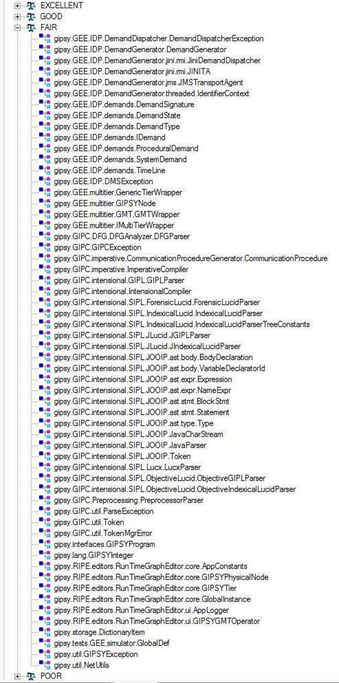}
\caption{GIPSY: Stability: Fair Classes}
\label{fig:50}

\end{figure}

\begin{figure} [ht!]
  \centering
    \includegraphics[width=0.5\textwidth]{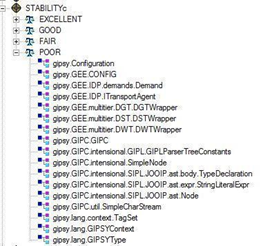}
\caption{GIPSY: Stability: Poor Classes}
\label{fig:51}

\end{figure}

\begin{figure} [ht!]
 \centering
  \includegraphics[width=0.5\textwidth]{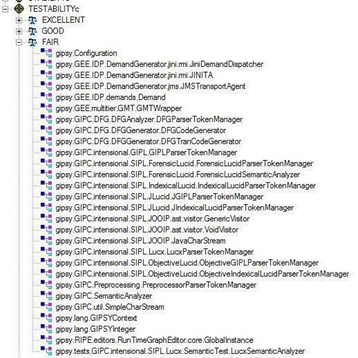}
\caption{GIPSY: Testability: Fair Classes}
\label{fig:52}

\end{figure}

\begin{figure} [ht!]
  \centering
    \includegraphics[width=0.5\textwidth]{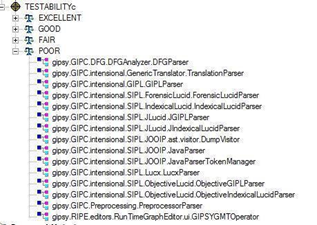}
\caption{GIPSY: Testability: Poor Classes}
\label{fig:53}

\end{figure}

\clearpage

\section{Appendix: B (MARF)}

\begin{figure} [ht!]
  \centering
    \includegraphics[width=0.5\textwidth]{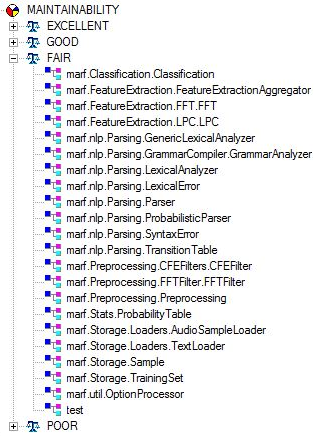}
\caption{MARF:Maintainability: Fair Classes}
\label{fig:57}

\end{figure}

\begin{figure} [ht!]
  \centering
    \includegraphics[width=0.5\textwidth]{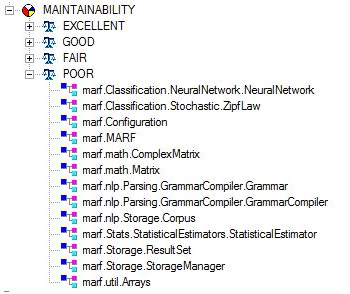}
\caption{MARF: Maintainability: Poor Classes}
\label{fig:58}

\end{figure}

\begin{figure} [ht!]
  \centering
    \includegraphics[width=0.5\textwidth]{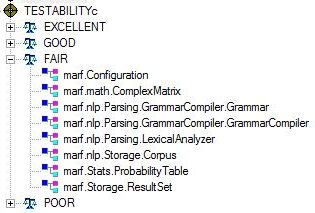}
\caption{MARF: Testability: Fair Classes}
 \label{fig:65}
\end{figure}

 \begin{figure} [ht!]
 \centering
\includegraphics[width=0.5\textwidth]{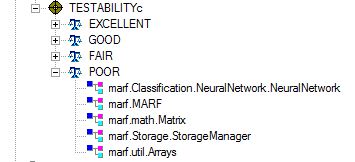}
\caption{MARF: Testability: Poor Classes}
\label{fig:66}
\end{figure}

\begin{figure} [ht!]
  \centering
    \includegraphics[width=0.5\textwidth]{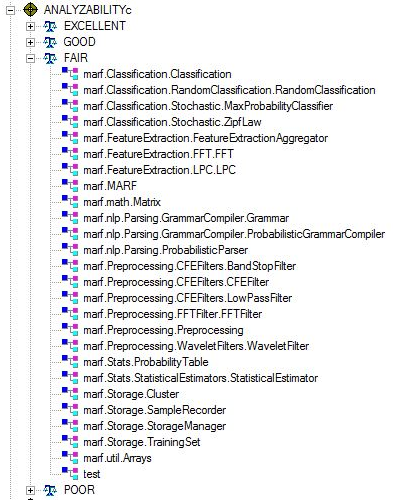}
\caption{MARF: Analyzability: Fair Classes}
\label{fig:59}

\end{figure}

\begin{figure} [ht!]
  \centering
    \includegraphics[width=0.5\textwidth]{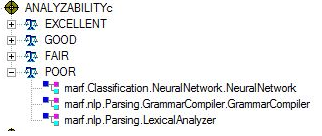}
\caption{MARF: Analyzability: Poor Classes}
\label{fig:60}

\end{figure}

\begin{figure} [ht!]
  \centering
    \includegraphics[width=0.5\textwidth]{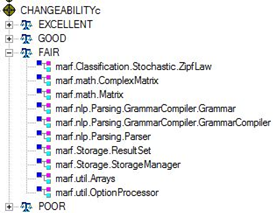}
\caption{MARF: Changeability: Fair Classes}
\label{fig:61}

\end{figure}

\begin{figure} [ht!]
  \centering
    \includegraphics[width=0.5\textwidth]{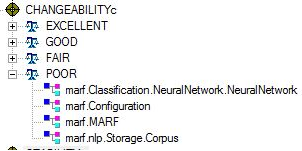}
\caption{MARF: Changeability: Poor Classes}
\label{fig:62}

\end{figure}

\pagebreak

\begin{figure} [ht!]
  \centering
    \includegraphics[width=0.5\textwidth]{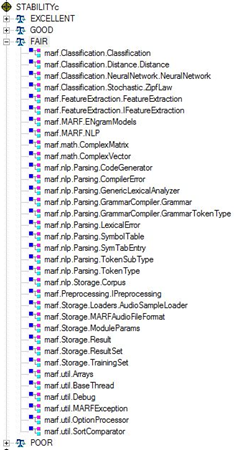}
\caption{MARF: Stability: Fair Classes}
\label{fig:63}

\end{figure}

\begin{figure} [ht!]
  \centering
    \includegraphics[width=0.5\textwidth]{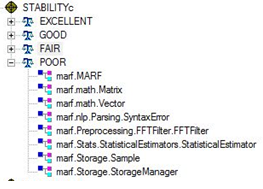}
\caption{MARF: Stability: Poor Classes}
 \label{fig:64}
\end{figure}

\end{document}